\documentclass[11pt]{article}

\usepackage[final]{acl}

\usepackage[T1]{fontenc}

\usepackage[utf8]{inputenc}
\usepackage{microtype}

\usepackage{times}
\usepackage{latexsym}
\usepackage{amsmath}
\usepackage{amssymb}
\usepackage{bbm}
\usepackage{makecell}
\usepackage{multirow}
\usepackage{array}
\usepackage{booktabs}
\usepackage{graphicx}
\usepackage{tcolorbox}
\usepackage{xcolor}
\usepackage{colortbl}
\usepackage{tabularx}
\usepackage{CJKutf8}

\newcommand{\zh}[1]{\begin{CJK*}{UTF8}{bsmi}#1\end{CJK*}}
\newcommand{\fr}[1]{#1}

\usepackage{algorithm}
\usepackage{algpseudocode}

\title{\textsc{Shift}: Mitigating Language Bias in Multilingual Information Retrieval}
\title{\textsc{Shift}: Semantic Harmonization via Index-side Feature Transformation for Multilingual Information Retrieval} 

\author{
    Youngjoon Jang, Seongtae Hong, Hyeonseok Moon\thanks{Corresponding authors}, Heuiseok Lim\footnotemark[1] \\
    Department of Computer Science and Engineering, Korea University \\
    \texttt{\{dew1701, ghdchlwls123, glee889, limhseok\}@korea.ac.kr}
}

\begin{document}
\maketitle
\begin{abstract}

With the rapid expansion of massive multilingual corpora, Multilingual Information Retrieval (MLIR) has emerged as a critical technology for global information access. MLIR enables users to retrieve semantically relevant documents from multilingual text collections using a single-language query. However, recent multilingual dense retrieval models often exhibit a strong preference for documents in the same language as the query. This leads to severe language bias, where top-ranked results are dominated by documents of specific languages, even when documents in other languages contain more semantically relevant information. To address this issue, we propose \textsc{Shift}, a training-free method applicable in the indexing stage. Specifically, \textsc{Shift} utilizes parallel translation pairs to estimate a relative language vector for each target language with respect to a source language. Subsequently, \textsc{Shift} corrects the language-specific offset by subtracting this relative language vector from document embeddings during indexing. Our comprehensive evaluation across four MLIR benchmarks and diverse dense retrieval models confirms that \textsc{Shift} can effectively mitigate language bias and enhance MLIR performance. The code for \textsc{Shift} is available at \url{https://github.com/yjoonjang/SHIFT}
\end{abstract}

\section{Introduction}
Information Retrieval (IR) constitutes the backbone of modern information systems, from traditional search engines to Retrieval-Augmented Generation (RAG)~\citep{ir1, ir2, rag1}. With the rapid expansion of massive multilingual corpora, Multilingual Information Retrieval (MLIR) has become essential for equitable information access. MLIR enables users to retrieve relevant information from multilingual text collections using a single-language query~\citep{mlir_notion}. In real-world applications, such as global enterprise search and multilingual RAG, robust MLIR is essential to supply LLMs with diverse multilingual knowledge, preventing them from generating answers based solely on particular languages.

Unlike Cross-Lingual Information Retrieval (CLIR), which assumes a one-to-one language mapping~\citep{clir1}, MLIR presents a more complex one-to-many challenge: simultaneously retrieving relevant documents across multiple target languages. Consequently, naive CLIR adaptations such as query translation, fail to perform effectively across massive multilingual corpora~\cite{transD1, transQ1}.
\begin{figure}[t!]
\centering
\includegraphics[width=1.0\linewidth]{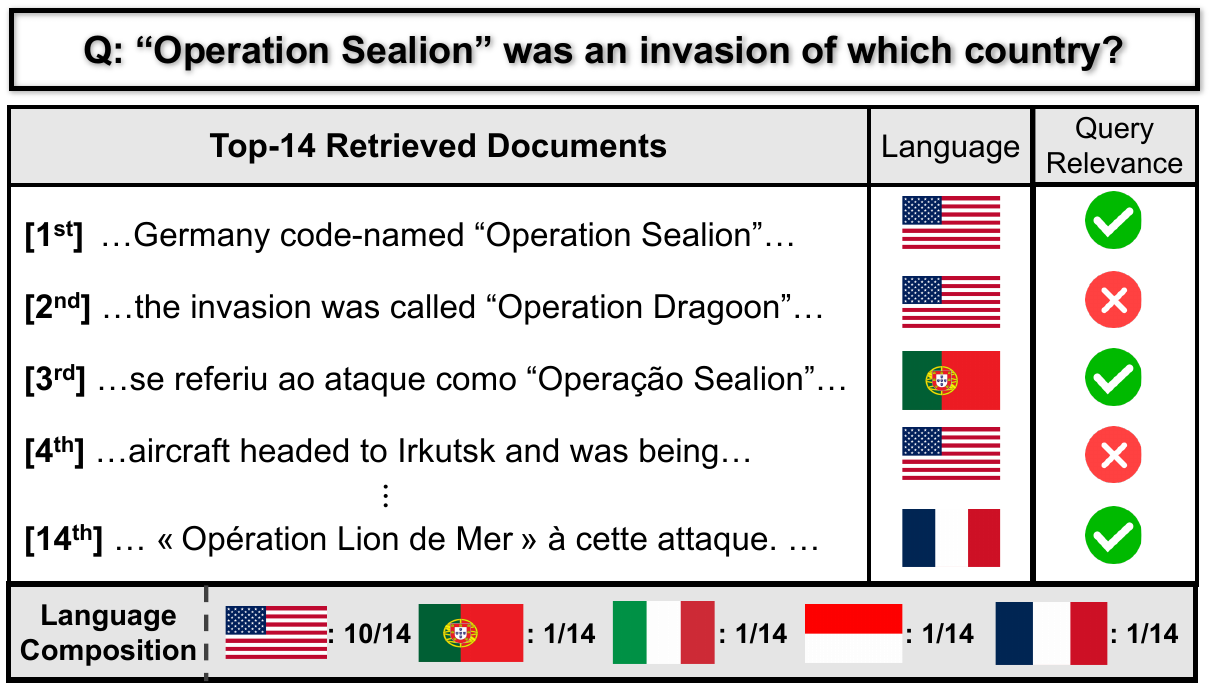}
\caption{An illustration of language bias in MLIR using \textit{multilingual-e5-large} on the Belebele dataset. The symbols $\checkmark$ and $\times$ denote semantic relevance. Despite the existence of semantically equivalent ground-truth documents across 14 languages, the top retrieved results are dominated by English documents (10/14).}
\label{fig:problem_statement}
\end{figure}

Ideally, an MLIR system should rank documents solely based on semantic similarity, invariant to linguistic surface forms. However, state-of-the-art multilingual dense retrievers suffer from severe language bias~\citep{bias1, bias2}. Rather than prioritizing semantic relevance, these models exhibit a disproportionate preference for documents written in the query language~\citep{park2025investigatinglanguagepreferencemultilingual, sharma-etal-2025-faux}.
As illustrated in Figure~\ref{fig:problem_statement}, retrieving with \textit{multilingual-e5-large} on the Belebele dataset reveals a pronounced language skew. Despite the existence of semantically equivalent ground-truth 
documents across 14 languages, the top ranks for an English query are saturated with semantically irrelevant English documents (10 out of 14). This indicates that language identity dominates semantic signals in the embedding space, distorting MLIR ranking behavior.

Conventional metrics often mask this bias by predominantly rewarding source-language retrievals. To explicitly quantify this, we introduce Target-Languages Recall@k (TLR@k), a novel metric measuring the retrieval effectiveness of non-query language documents. Furthermore, to mitigate this bias, we propose \textsc{Shift}, a training-free method applied during the indexing stage. By estimating a relative language vector from parallel translation pairs and subtracting it from document embeddings, \textsc{Shift} corrects the language-specific offset and forces the model to rely purely on underlying semantic signals. In our experiments, \textsc{Shift} yields consistent improvements, achieving up to a 16.4\% relative gain in TLR@20.
The main contributions of this work are as follows:

\begin{itemize}
    \item We demonstrate that \textsc{Shift} is a universally effective solution, consistently improving MLIR performance and mitigating language bias across diverse dense retrieval models.
    \item We introduce Target-Languages Recall@k (TLR@k), a novel metric that explicitly quantifies language bias by measuring the retrieval of non-query language documents.
    \item Through detailed analysis and experiments, we provide empirical evidence that \textsc{Shift} successfully realigns multilingual documents based on semantics, correcting the skewed geometry of the embedding space.
\end{itemize}

\section{Preliminaries}

\subsection{Related Works}
The concept of "multilingual" in Information Retrieval (IR) can be interpreted in various contexts. For instance, multilingual IR is often referred to as monolingual retrieval across multiple languages~\citep{hull1996querying, blloshmi2021ir} or as CLIR tasks involving multiple languages~\citep{10.1007/3-540-45691-0_2, braschler2002clef, lawrie2022hc4, mitamura2008overview}. In this work, however, we adopt the definition of MLIR established by the Cross-Language Evaluation Forum (CLEF): a task where a user retrieves information from a mixed-language document collection using a query in a single language~\citep{peters2002importance}.

Traditional MLIR approaches relied on Machine Translation (MT) to translate either the entire document collection or the query~\citep{darwish2003probabilistic, kraaij2003embedding, mcnamee2002comparing, transD1, transD2, transQ1, transQ2, transQ3}. However, these methods incur significant computational costs or introduce systematic ranking bias due to disparate score distributions~\citep{huang2023soft, lawrie2023neural}.

Recently, dense retrieval models based on multilingual Pre-trained Language Models (mPLMs) have become a standard, demonstrating robust performance in MLIR tasks~\citep{conneau-etal-2020-unsupervised, chen-etal-2024-m3, yu2024arcticembed20multilingualretrieval}. In MLIR scenarios, dense retrievers encode both queries and documents into a shared embedding space, enabling efficient retrieval via Approximate Nearest Neighbor (ANN) search using scoring functions such as dot product or cosine similarity~\citep{ann1, ann2, ann3}.

\subsection{Language Bias in MLIR}
However, the embedding spaces of dense retrievers fail to achieve perfect linguistic alignment. Consequently, documents representing identical meanings but written in different languages are often mapped to distant locations, leading to language bias that distorts the ranking~\citep{bias1, bias4}. An ideal MLIR system should rank documents solely based on semantic similarity, independent of the document's language. Yet, in practice, a recurring phenomenon is observed where documents in the same language as the query disproportionately occupy the top ranks~\citep{ park2025investigatinglanguagepreferencemultilingual, sharma-etal-2025-faux}. This bias increases the exposure of specific languages within multilingual collections, simultaneously undermining the fairness of information access and user experience.

To mitigate this issue, prior research has largely adopted two approaches. The first involves strengthening the model's linguistic alignment through additional training~\citep{hu2023language, bias1, huang2023soft, Yang_2024}. 
While effective, this approach requires high computational costs due to large-scale training and necessitates retraining whenever the language combination of the model or data changes. 
The second approach addresses language bias at the pipeline level without modifying the model. 
A representative line of work translates each query into multiple languages and fuses language-specific ranked lists~\citep{lawrie2025overviewtrec2024neuclir, yang2025hltcoetrec2024neuclir}. 
Although effective, such pipelines typically increase system complexity and inference-time latency.
Another line mitigates bias by post-hoc normalization of multilingual embeddings (e.g., language-wise centering~\citep{libovicky-etal-2020-language} or projection-based methods~\citep{langdetect1}) without any additional training.
While these strategies can improve retrieval performance, they may inadvertently suppress semantic signal when enforcing language-agnostic representations, and often require additional query-side processing, increasing inference-time latency.

\section{\textsc{Shift}}
In this work, we propose \textsc{Shift}, a training-free method designed to align multilingual representations at the indexing stage.
Unlike approaches that necessitate computationally expensive training or impose additional inference latency, \textsc{Shift} employs a linear displacement strategy to correct spatial misalignment.
This method preserves the high-dimensional structure of the original embeddings to the greatest extent possible, thereby improving MLIR performance with zero additional query-time overhead.
\subsection{Estimating Relative Language Vectors}
Let $\mathcal{L}_{\mathrm{tgt}}$ denote the set of target languages, and let $\ell_{\mathrm{tgt}}$ be an element of $\mathcal{L}_{\mathrm{tgt}}$.
The core mechanism of \textsc{Shift} is to estimate a relative language vector $V_{\ell_{\mathrm{tgt}}}$ for each target language, with respect to a source language $\ell_{\mathrm{src}}$.
To estimate these vectors, we utilize parallel documents consisting of source--target translation pairs from mMARCO~\citep{bonifacio2022mmarcomultilingualversionms}.
For each target language $\ell_{\mathrm{tgt}}$, we collect a set of translation pairs:
\begin{equation*}
\mathcal{P}_{\ell_{\mathrm{tgt}}}
= \big\{ (D_i^{\ell_{\mathrm{src}}},\, D_i^{\ell_{\mathrm{tgt}}}) \big\}_{i=1}^{N},
\quad \ell_{\mathrm{tgt}} \in \mathcal{L}_{\mathrm{tgt}} 
\end{equation*}
Each pair consists of a source-language document $D_i^{\ell_{\mathrm{src}}}$ and its translation
$D_i^{\ell_{\mathrm{tgt}}}$, sharing identical semantic meaning.
We first encode each document into the embedding space using the dense retrieval model $f(\cdot)$:
\begin{equation*}
    z_i^{\ell_{\mathrm{src}}} = f(D_i^{\ell_{\mathrm{src}}}), \quad
    z_i^{\ell_{\mathrm{tgt}}} = f(D_i^{\ell_{\mathrm{tgt}}})
\end{equation*}
The difference between the two embeddings,
$\Delta_i^{\ell_{\mathrm{tgt}}} = z_i^{\ell_{\mathrm{tgt}}} - z_i^{\ell_{\mathrm{src}}}$,
approximates the representational offset arising solely from the language shift from
$\ell_{\mathrm{src}}$ to $\ell_{\mathrm{tgt}}$ within two semantically equivalent documents.
We define the relative language vector of each target language $\ell_{\mathrm{tgt}}$ with respect to the source language
by averaging this offset over all $N$ translation pairs:
\begin{equation}
    V_{\ell_{\mathrm{tgt}}}
    = \frac{1}{N} \sum_{i=1}^{N}
      \big( z_i^{\ell_{\mathrm{tgt}}} - z_i^{\ell_{\mathrm{src}}}),
    \quad \forall \ell_{\mathrm{tgt}} \in \mathcal{L}_{\mathrm{tgt}}
\label{eq:lang_estimation}
\end{equation}


\subsection{Index-side Language Shift}
\label{sec:index_side_language_shift}
After estimating the relative language vectors, we calibrate document embeddings during indexing via linear subtraction.
Specifically, for each document pre-labeled with its language $\ell$, we subtract the corresponding vector
if $\ell \neq \ell_{\mathrm{src}}$.
This adjustment is controlled by a hyperparameter $\alpha$, which scales the transformation magnitude: larger $\alpha$ subtracts more of $V_{\ell_{\mathrm{tgt}}}$, pulling target-language document embeddings closer to the source-language embedding space. 
We use a single global $\alpha$ shared by all documents and sweep $\alpha \in \{0.1,0.2,\ldots,1.0\}$ to report sensitivity to $\alpha$.
We do not tune $\alpha$ per dataset, language, or query set.
The overall indexing procedure is summarized in Algorithm~\ref{alg:shift}.

\begin{algorithm}[t]
\caption{\textsc{Shift} Indexing}
\label{alg:shift}
\begin{algorithmic}[1]
    \Require
        Labeled Documents $\mathcal{D} = \{(d_j, \ell_j)\}_{j=1}^{M}$,
        Model $f$,
        Source Language $\ell_{\mathrm{src}}$,
        Target Language Set $\mathcal{L}_{\mathrm{tgt}}$,
        Relative Language Vectors $\{V_{\ell}\}_{\ell \in \mathcal{L}_{\mathrm{tgt}}}$,
        Scale factor $\alpha \in [0, 1]$
    \Statex
    \For{each $(d, \ell) \in \mathcal{D}$}
        \State $z \leftarrow f(d)$
        \If{$\ell = \ell_{\mathrm{src}}$}
            \State $\tilde{z} \leftarrow z$
        \Else
            \State $\tilde{z} \leftarrow z - \alpha \, V_{\ell}$
        \EndIf
        \State Add $\tilde{z}$ to the index
    \EndFor
\end{algorithmic}
\end{algorithm}

\subsection{Retrieval}
The retrieval process on the constructed index $\tilde{\mathcal{Z}} = \{\tilde{z}_1, \dots, \tilde{z}_M\}$
is performed identically to standard dense retrieval procedures.
Given a query $q$ in the source language, we generate the query embedding
$z_q = f(q)$ using the same embedding model $f$ used for indexing.
The final retrieval score is calculated as the similarity between the query embedding and the calibrated document embedding $\tilde{z}$ in the index:
\begin{equation*}
    \text{score}(q, d) = \text{sim}(z_q, \tilde{z})
\end{equation*}
All models used in our experiments utilize cosine similarity for $\text{sim}(\cdot, \cdot)$.
Since the documents processed with \textsc{Shift} are already aligned to the source language space,
no additional transformation is required for the query. 
Consequently, our method introduces no additional latency at query-time compared to baseline models.

\section{Experimental Setup}

\subsection{Models \& Resources}
\label{sec:setup_models_resources}

\paragraph{Relative Language Vector Estimation}
To estimate relative language vectors, we use the mMARCO dataset~\citep{bonifacio2022mmarcomultilingualversionms}.
mMARCO provides machine-translated versions of the original English content in 14 languages (via Google Translate), yielding aligned text pairs across languages.
We leverage all 533k aligned passage pairs for language vector estimation.

\paragraph{Multilingual Retrievers}
To verify the generalizability of \textsc{Shift}, we employ a diverse set of multilingual dense retrievers featuring different architectures. We include encoder-based models such as \textit{embeddinggemma-300m}~\citep{vera2025embeddinggemma}, \textit{multilingual-e5-large}~\citep{me5}, and \textit{bge-m3}~\citep{bge-m3}, as well as decoder-based models including \textit{Qwen3-Embedding-0.6B}~\cite{qwen3embedding}, \textit{llama-nemotron-embed-1b-v2}~\citep{babakhin2025llamaembednemotron8buniversaltextembedding}, and \textit{gte-Qwen2-1.5B-instruct}~\citep{mgte}. Implementation details are provided in Appendix~\ref{sec:implementation_details}.

\subsection{Evaluation Datasets}
\label{sec:setup_datasets}
To rigorously evaluate the robustness of \textsc{Shift}, we employ four benchmarks adapted to the MLIR setting.
In our main experiments (Section~\ref{sec:results_main}), we use English queries as the source language and retrieve from a multilingual corpus.
We further evaluate Chinese, Vietnamese, and Hindi queries as source languages in Section~\ref{sec:results_otherlangs}.
Below, we briefly describe each dataset and our MLIR construction procedure.

\paragraph{Belebele \& XQuAD}
Belebele~\citep{belebele} is a fully parallel 122-language MRC benchmark built on the FLORES-200 translation benchmark~\citep{nllbteam2022languageleftbehindscaling}, and XQuAD~\citep{xquad} is a fully parallel 10-language translation of a subset of SQuAD v1.1~\citep{rajpurkar-etal-2016-squad}.
To adapt these datasets for MLIR, we construct a multilingual retrieval including all language versions of the context passages, inspired by the multilingual pool construction in LAReQA~\citep{lareqa}. Accordingly, for each source query, we treat the source-language gold context and all target-language versions of that context as positive documents.
In our experiments, we utilize the subset of languages that overlap with mMARCO, resulting in 14 languages for Belebele and 8 languages for XQuAD.

\paragraph{MLQA}
MLQA~\citep{mlqa} is an extractive QA benchmark covering 7 languages. Unlike fully parallel datasets where every item exists in all languages, MLQA is constructed such that each instance is aligned across a subset of languages (typically four). For the MLIR setup, we aggregate contexts from all 7 languages in the document collection. For a given source query, all available contexts across different languages are treated as positive documents. This implies that the number of positive documents and their language composition may vary per query, but all semantically linked multilingual contexts are considered as positives.

\paragraph{MultiEuP-v2}
MultiEuP-v2~\citep{bias1} is a dataset derived from European Parliament proceedings, spanning 24 languages, using debate titles as queries and individual speeches as documents. 
We configure the MLIR task such that for a single source query (debate title), all speeches belonging to that debate topic are considered positive documents. Thus, the positive document includes both the speech in the original language and speeches in other languages within the same topic.
Consistent with our training setup, we evaluate on the 6 languages that overlap with mMARCO.

We summarize the dataset statistics in Table~\ref{tab:dataset_stats}, with additional details in Appendix~\ref{sec:evaluation_dataset_details}.
\begin{table}[h!]
\centering
\normalsize
\resizebox{1.0\linewidth}{!}
{
\begin{tabular}{l|cccc}
\toprule
\textbf{Data Statistics} & \textbf{Belebele} & \textbf{MLQA} & \textbf{XQuAD} & \textbf{MultiEup-v2} \\
\midrule
\# of Languages & \makecell{14} & \makecell{7} & \makecell{8} & \makecell{6} \\
Fully Parallel & O & X & O & X \\
\midrule
\# of Queries & 900 & 11,582 & 1,190 & 1,781 \\
\# of Documents & 6,832 & 36,799 & 1,920 & 35,444 \\
\midrule
\# of $d^+/q$ & 14 & 3.65 & 8 & 19.9 \\
\# of $d^+_{en}/q$ & 1 & 1 & 1 & 7.91 \\
\bottomrule
\end{tabular}
}
\caption{Statistics of the MLIR benchmark datasets. \# of $d^{+}/q$ denotes the  number of positive documents per query, and \# of $d_{en}^{+}/q$ indicates the number of English positive documents per query.}
\label{tab:dataset_stats}
\end{table}

\begin{table*}[t!]
\centering
\resizebox{1.0\textwidth}{!}{%
\begin{tabular}{cl|cccccccc||cc}
\toprule
\multirow{2}{*}{\textbf{Archi}} & \multicolumn{1}{c|}{\multirow{2}{*}{\textbf{Model}}} & \multicolumn{2}{c}{\textbf{Belebele}} & \multicolumn{2}{c}{\textbf{MLQA}} & \multicolumn{2}{c}{\textbf{XQuAD}} & \multicolumn{2}{c||}{\textbf{MultiEup-v2}} & \multicolumn{2}{c}{\textbf{AVG}} \\ 
\cmidrule(lr){3-4} \cmidrule(lr){5-6} \cmidrule(lr){7-8} \cmidrule(lr){9-10} \cmidrule(lr){11-12} 
 & \multicolumn{1}{c|}{} & nDCG & Recall & nDCG & Recall & nDCG & Recall & nDCG & Recall & nDCG & Recall \\ \midrule
 
\multirow{6}{*}{\rotatebox[origin=c]{90}{\textbf{encoder}}} 
  & embeddinggemma-300m & 0.924 & 0.926 & 0.664 & 0.740 & 0.947 & 0.970 & 0.495 & 0.433 & 0.758 & 0.767 \\
  & + \textsc{Shift} & \textbf{0.926} & \textbf{0.928} & \textbf{0.671} & \textbf{0.748} & \textbf{0.951} & \textbf{0.972} & \textbf{0.511} & \textbf{0.443} & \textbf{0.765} & \textbf{0.773} \\ \cmidrule{2-12}
 
 & multilingual-e5-large & 0.816 & 0.812 & 0.494 & 0.523 & 0.855 & 0.900 & 0.367 & 0.302 & 0.633 & 0.634 \\
 & + \textsc{Shift} & \textbf{0.910} & \textbf{0.907} & \textbf{0.649} & \textbf{0.721} & \textbf{0.944} & \textbf{0.963} & \textbf{0.442} & \textbf{0.375} & \textbf{0.737} & \textbf{0.742} \\ \cmidrule{2-12} 
 
 & bge-m3 & 0.874 & 0.871 & 0.607 & 0.679 & 0.923 & 0.947 & 0.455 & 0.387 & 0.715 & 0.721 \\
 & + \textsc{Shift} & \textbf{0.876} & \textbf{0.874} & \textbf{0.609} & \textbf{0.682} & \textbf{0.925} & \textbf{0.950} & \textbf{0.460} & \textbf{0.391} & \textbf{0.717} & \textbf{0.724} \\ \midrule
 
\multirow{6}{*}{\rotatebox[origin=c]{90}{\textbf{decoder}}} 
 & Qwen3-Embedding-0.6B & 0.890 & 0.887 & 0.611 & 0.688 & 0.914 & 0.940 & 0.468 & 0.403 & 0.721 & 0.729 \\
 & + \textsc{Shift} & \textbf{0.895} & \textbf{0.892} & 0.611 & 0.688 & \textbf{0.916} & \textbf{0.942} & \textbf{0.476} & \textbf{0.411} & \textbf{0.725} & \textbf{0.733} \\ \cmidrule{2-12} 
 
 & llama-nemotron-embed-1b-v2 & 0.688 & 0.681 & 0.523 & 0.602 & 0.927 & 0.957 & 0.292 & 0.251 & 0.608 & 0.623 \\
 & + \textsc{Shift} & \textbf{0.700} & \textbf{0.697} & \textbf{0.527} & \textbf{0.609} & \textbf{0.931} & \textbf{0.960} & \textbf{0.312} & \textbf{0.273} & \textbf{0.618} & \textbf{0.635} \\ \cmidrule{2-12} 
 
 & gte-Qwen2-1.5B-instruct & 0.904 & 0.901 & 0.621 & 0.691 & 0.926 & 0.946 & 0.496 & 0.430 & 0.737 & 0.742 \\
 & + \textsc{Shift} & \textbf{0.914} & \textbf{0.912} & \textbf{0.639} & \textbf{0.716} & \textbf{0.936} & \textbf{0.955} & \textbf{0.519} & \textbf{0.451} & \textbf{0.752} & \textbf{0.758} \\ \bottomrule
\end{tabular}
}
\caption{Top-20 retrieval performance (nDCG@20, Recall@20) of baseline models and \textsc{Shift} across four MLIR benchmarks, evaluated with English as the source query language. 
We conduct a global grid search over $\alpha \in {0.1,\dots,1.0}$ and report the best $\alpha$ for each model in subsequent experiments, as described in Section~\ref{sec:analysis_alpha}. 
}
\label{tab:retrieval_performance}
\end{table*}

\subsection{Evaluation Metrics}
\label{sec:setup_metrics}
We employ standard metrics, nDCG@20 and Recall@20, to measure retrieval accuracy. 
However, these conventional metrics do not reveal whether relevant documents beyond the query language are retrieved. In MLIR benchmarks such as MultiEuP-v2, for instance, a model can obtain high nDCG and Recall scores by disproportionately retrieving source-language documents, despite the presence of relevant documents in other languages.


To better capture this aspect, we introduce Target-Languages Recall@k (TLR@k).
Essentially, this metric computes the recall score exclusively on the subset of relevant documents written in languages other than the query language.
For a query $q$, let $\mathcal{D}_q^{+}$ denote the set of relevant documents to the query.
We define the subset of relevant documents in target languages as
$R_q^{\mathrm{tgt}} = \{ d \in \mathcal{D}_q^{+} \mid L(d) \neq L(q) \}$.
TLR@k is then calculated as the average recall of these documents:
\begin{equation*}
\mathrm{TLR@k}
= \frac{1}{|\mathcal{Q}|}
  \sum_{q \in \mathcal{Q}}
  \frac{\bigl|\mathrm{top}_k(q) \cap R_q^{\mathrm{tgt}}\bigr|}
       {\bigl|R_q^{\mathrm{tgt}}\bigr|}
\end{equation*}
A higher TLR@k indicates that the model successfully retrieves target-language documents, complementing conventional accuracy metrics that may be dominated by source-language documents.

\section{Results \& Analysis}

\subsection{Main Results}
\label{sec:results_main}

\paragraph{Retrieval Performance}
Table~\ref{tab:retrieval_performance} presents the retrieval performance before and after applying \textsc{Shift} across four MLIR benchmarks.
The results demonstrate that \textsc{Shift} consistently achieves significant performance gains across all settings.
For the encoder-based model, nDCG@20 for the \textit{embeddinggemma-300m} model increased from 0.758 to 0.765 on average, while it significantly rose from 0.633 to 0.737 for \textit{multilingual-e5-large} after applying \textsc{Shift}.
This improvement extends to decoder-based models,
 suggesting that \textsc{Shift} effectively mitigates the representational discrepancies between source queries and multilingual documents across architectures and model scales.

\begin{table}[h!]
\centering
\normalsize
\resizebox{1.0\linewidth}{!}{
\begin{tabular}{l|cccc||c}
\toprule
\multicolumn{1}{c|}{\textbf{Model}} & \textbf{Belebele} & \textbf{MLQA} & \textbf{XQuAD} & \textbf{MultiEup-v2} & \textbf{AVG} \\ \midrule

 embeddinggemma & 0.922 & 0.706 & 0.966 & 0.318 & 0.728 \\
 + \textsc{Shift} & \textbf{0.926} & \textbf{0.736} & \textbf{0.971} & \textbf{0.371} & \textbf{0.751} \\ \cmidrule{1-6}
 
 mE5-large & 0.798 & 0.380 & 0.886 & 0.095 & 0.540 \\
 + \textsc{Shift} & \textbf{0.898} & \textbf{0.654} & \textbf{0.956} & \textbf{0.268} & \textbf{0.694} \\ \cmidrule{1-6}
 
 bge-m3 & 0.864 & 0.637 & 0.942 & 0.293 & 0.684 \\
 + \textsc{Shift} & \textbf{0.868} & \textbf{0.643} & \textbf{0.945} & \textbf{0.300} & \textbf{0.689} \\ \midrule

 Qwen3 & 0.880 & 0.650 & 0.933 & 0.302 & 0.691 \\
 + \textsc{Shift} & \textbf{0.885} & \textbf{0.651} & \textbf{0.934} & \textbf{0.322} & \textbf{0.698} \\ \cmidrule{1-6}
 
 llama-nemotron & 0.667 & \textbf{0.554} & 0.951 & 0.123 & 0.574 \\
 + \textsc{Shift} & \textbf{0.683} & 0.553 & \textbf{0.953} & \textbf{0.156} & \textbf{0.586} \\ \cmidrule{1-6}
 
 gte-Qwen2 & 0.893 & 0.629 & 0.939 & 0.303 & 0.691 \\
 + \textsc{Shift} & \textbf{0.908} & \textbf{0.683} & \textbf{0.951} & \textbf{0.378} & \textbf{0.730} \\ \bottomrule
\end{tabular}
}
\caption{Comparison of TLR@20 across different models and datasets.}
\label{tab:tlr}
\end{table}

\paragraph{Mitigation of Language Bias} 
To verify that the performance improvement stems from increased exposure of relevant target-language documents rather than merely higher rankings of source-language documents, we analyze the TLR@20 metric in Table~\ref{tab:tlr}.
Notably, standard metrics such as Recall can mask underlying language bias, as a model may achieve high scores solely by retrieving documents in the source language.
For instance, on MultiEup-v2, \textit{multilingual-e5-large} attains a reasonable Recall@20 of 0.302 in Table~\ref{tab:retrieval_performance}, yet its TLR@20 is only 0.095 in Table~\ref{tab:tlr}.
This gap suggests that the model predominantly retrieves relevant documents in the source language, under-exposing relevant target-language documents.
Applying \textsc{Shift} alleviates this issue, yielding a substantial increase in TLR@20 across all models.
This confirms that \textsc{Shift} bridges the representation gap between the query and document languages, effectively promoting relevant target-language documents to the top ranks. We provide additional comparisons to representative debiasing baselines in Appendix~\ref{sec:baseline}.

\begin{table*}[t]
\centering
\resizebox{\textwidth}{!}{%
\begin{tabular}{cl|cc|cc|cc|cc|cc|cc|cc|cc}
\toprule
\multirow{3}{*}{\textbf{Query}} & \multirow{3}{*}{\textbf{Dataset}} & \multicolumn{4}{c|}{\textbf{multilingual-e5-large}} & \multicolumn{4}{c|}{\textbf{bge-m3}} & \multicolumn{4}{c|}{\textbf{llama-nemotron-embed-1b-v2}} & \multicolumn{4}{c}{\textbf{gte-Qwen2-1.5B-instruct}} \\ \cmidrule{3-18} 
 &  & \multicolumn{2}{c}{\textbf{nDCG@20}} & \multicolumn{2}{c|}{\textbf{TLR@20}} & \multicolumn{2}{c}{\textbf{nDCG@20}} & \multicolumn{2}{c|}{\textbf{TLR@20}} & \multicolumn{2}{c}{\textbf{nDCG@20}} & \multicolumn{2}{c|}{\textbf{TLR@20}} & \multicolumn{2}{c}{\textbf{nDCG@20}} & \multicolumn{2}{c}{\textbf{TLR@20}} \\ \cmidrule{3-18} 
 &  & Base & \textsc{Shift} & Base & \textsc{Shift} & Base & \textsc{Shift} & Base & \textsc{Shift} & Base & \textsc{Shift} & Base & \textsc{Shift} & Base & \textsc{Shift} & Base & \textsc{Shift} \\ \midrule
\multirow{3}{*}{ZH} & Belebele & 0.321 & \textbf{0.825} & 0.195 & \textbf{0.802} & 0.824 & \textbf{0.831} & 0.813 & \textbf{0.818} & 0.597 & \textbf{0.661} & 0.572 & \textbf{0.654} & 0.875 & \textbf{0.884} & 0.867 & \textbf{0.878} \\
 & MLQA & 0.317 & \textbf{0.549} & 0.059 & \textbf{0.483} & 0.599 & \textbf{0.607} & 0.637 & \textbf{0.653} & 0.483 & \textbf{0.519} & 0.487 & \textbf{0.552} & 0.598 & \textbf{0.623} & 0.633 & \textbf{0.673} \\
 & XQuAD & 0.457 & \textbf{0.880} & 0.383 & \textbf{0.907} & 0.890 & \textbf{0.898} & 0.926 & \textbf{0.931} & 0.870 & \textbf{0.889} & 0.917 & \textbf{0.930} & 0.903 & \textbf{0.916} & 0.935 & \textbf{0.942} \\ \midrule
\multirow{3}{*}{VI} & Belebele & 0.596 & \textbf{0.866} & 0.551 & \textbf{0.853} & 0.836 & \textbf{0.839} & 0.828 & \textbf{0.831} & 0.461 & \textbf{0.605} & 0.421 & \textbf{0.581} & 0.658 & \textbf{0.851} & 0.629 & \textbf{0.842} \\
 & MLQA & 0.416 & \textbf{0.630} & 0.272 & \textbf{0.674} & 0.606 & \textbf{0.610} & 0.652 & \textbf{0.663} & 0.427 & \textbf{0.494} & 0.402 & \textbf{0.535} & 0.376 & \textbf{0.543} & 0.259 & \textbf{0.572} \\
 & XQuAD & 0.689 & \textbf{0.911} & 0.717 & \textbf{0.941} & 0.890 & \textbf{0.896} & 0.920 & \textbf{0.926} & 0.794 & \textbf{0.870} & 0.832 & \textbf{0.905} & 0.673 & \textbf{0.866} & 0.702 & \textbf{0.897} \\ \midrule
\multirow{3}{*}{HI} & Belebele & 0.534 & \textbf{0.808} & 0.482 & \textbf{0.789} & 0.775 & \textbf{0.780} & 0.764 & \textbf{0.771} & 0.629 & \textbf{0.668} & 0.611 & \textbf{0.657} & 0.512 & \textbf{0.643} & 0.466 & \textbf{0.613} \\
 & MLQA & 0.414 & \textbf{0.614} & 0.275 & \textbf{0.646} & 0.591 & \textbf{0.595} & 0.641 & \textbf{0.649} & 0.493 & \textbf{0.528} & 0.522 & \textbf{0.588} & 0.352 & \textbf{0.458} & 0.262 & \textbf{0.483} \\
 & XQuAD & 0.697 & \textbf{0.902} & 0.722 & \textbf{0.927} & 0.882 & \textbf{0.887} & 0.908 & \textbf{0.913} & 0.859 & \textbf{0.877} & 0.904 & \textbf{0.921} & 0.681 & \textbf{0.799} & 0.704 & \textbf{0.838} \\ \bottomrule
\end{tabular}%
}
\caption{Top-20 retrieval performance (nDCG@20, TLR@20) using non-English languages (Chinese, Vietnamese, Hindi) as the source query language.}
\label{tab:retrieval_multilingual_queries}
\end{table*}



\subsection{Language Distribution Analysis}
\label{sec:analysis_distribution}
\begin{figure}[t!]
\centering
\includegraphics[width=1.0\linewidth]{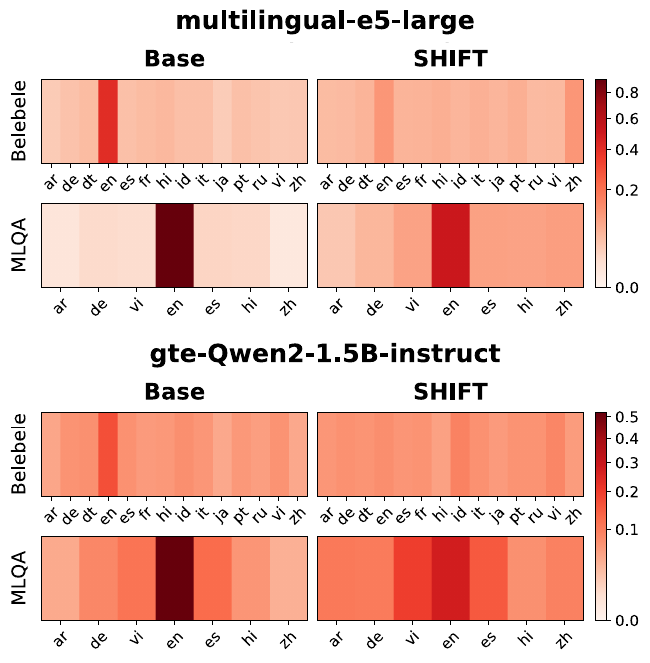}
 \caption{Visualization of language distribution in the top-20 retrieved documents for English queries. The heatmap intensity represents the density of retrieved documents for each language, where darker shades indicate a higher concentration of documents. Distributions for additional models and data are provided in Appendix~\ref{sec:appen_heatmap}.} 
 \label{fig:heatmap}
\end{figure}

To complement the TLR analysis, Figure~\ref{fig:heatmap} visualizes the language distribution of the top-20 documents retrieved by \textit{multilingual-e5-large} and \textit{gte-Qwen2-1.5B-instruct}.
The heatmap compares the density of retrieved documents by language before and after applying \textsc{Shift}, and thereby provides a qualitative view of the language composition and diversity of the retrieved results.

Our experimental results demonstrate that existing baseline models have a certain language preference.
Given an English source query, the models assign disproportionately high similarity scores to English documents, filling the top ranks with English content. 
Conversely, documents in target languages are ranked lower, appearing as inactive or very faint regions on the heatmap.
This indicates that 
the model's similarity space favors certain language regions even when relevant documents exist more broadly in target languages.

After applying \textsc{Shift}, this concentration becomes noticeably less pronounced.
The density previously concentrated in English becomes more evenly dispersed across languages, and previously inactive regions become clearly visible.
This suggests that \textsc{Shift} encourages stronger cross-lingual alignment between the source-language query and target-language documents, enabling the retriever to return a markedly more balanced language composition among the top-ranked results.
Importantly, this qualitative trend is consistent with the TLR@20 improvements in Table~\ref{tab:tlr}, suggesting that higher TLR@20 coincides with reduced query-language dominance and increased presence of other languages in the top ranks.

\subsection{\textsc{Shift} in Multi-Source Languages}
\label{sec:results_otherlangs}
To demonstrate the universality of \textsc{Shift} and verify that it functions effectively regardless of the source language, we conduct additional experiments using three different languages as new source queries.

\paragraph{Experimental Setup}
We select Chinese (zh), Vietnamese (vi), and Hindi (hi) as non-English source (query) languages.
These languages overlap with mMARCO and our evaluation benchmarks and cover distinct scripts and character distributions.
For each run, we treat the selected query language as the source language $\ell_{\mathrm{src}}$ and re-estimate the relative language vectors $V_{\ell_\mathrm{tgt}}$ using Eq.~\ref{eq:lang_estimation} for all target languages $\ell_{\mathrm{tgt}} \in \mathcal{L}_{\mathrm{tgt}}$.
Subsequently, during indexing (as in Section~\ref{sec:index_side_language_shift}), all documents are shifted into the embedding space of the $\ell_{\mathrm{src}}$.
For evaluation, we reprocess Belebele, MLQA, and XQuAD to construct MLIR test sets for each $\ell_{\mathrm{src}}$. We provide the constructed 
dataset statistics in Appendix~\ref{sec:multi_source_language_evaluation_dataset}.
We report nDCG@20 and TLR@20 to measure overall retrieval effectiveness and target-languages coverage.

\paragraph{Results}
Table~\ref{tab:retrieval_multilingual_queries} presents retrieval performance when non-English languages are set as the source query.
The results show common performance improvements across all source languages and models upon applying \textsc{Shift}, proving that our method is not limited to a specific source language.
The most notable change is observed in the \textit{multilingual-e5-large} model; 
it records significant increases in both nDCG@20 and TLR@20 compared to the base model, demonstrating the method's effectiveness.

Another noteworthy observation is the \textit{bge-m3} model. 
As shown in Table~\ref{tab:retrieval_performance}, when the source language is English, the baseline performance is already strong and the gains from \textsc{Shift} are marginal.
However, when non-English languages serve as source languages, applying \textsc{Shift} substantially improves TLR@20 on MLQA.
This suggests that while state-of-the-art models like \textit{bge-m3} are strongly aligned around English, their alignment is relatively less consistent when the source language is non-English.
Consequently, \textsc{Shift} effectively corrects these non-English-centric alignment discrepancies, demonstrating robustness across different source languages.
To reflect more realistic scenarios where users query in various languages, we provide additional experiments on Multilingual-to-Multilingual Information Retrieval (M2MIR) in Appendix~\ref{sec:m2mir}.

\subsection{Analysis of Scale Factor $\alpha$}
\label{sec:analysis_alpha}
The scale factor $\alpha$ is a key hyperparameter in \textsc{Shift} that controls the strength of the language-vector subtraction. 
Varying $\alpha$ adjusts the degree of debiasing applied to the embedding space, and an overly aggressive subtraction may lead to over-correction for some models.
To study this effect, we conduct a grid search over $\alpha$ from 0.1 to 1.0 and report the average nDCG@20 across all datasets.

\begin{figure}[t!]
\centering
\includegraphics[width=1.0\linewidth]{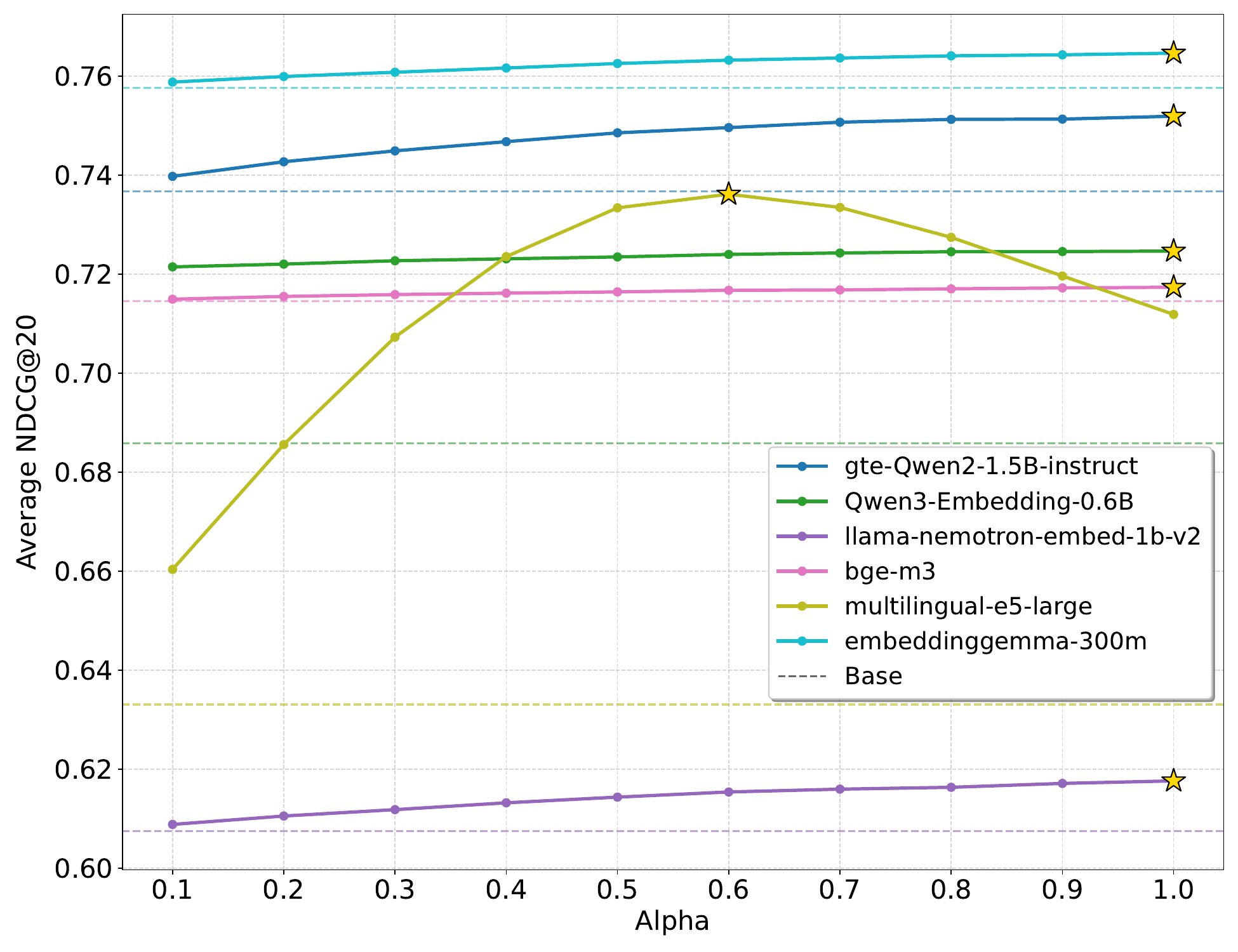}
\caption{Impact of the scale factor $\alpha$ on the average nDCG@20 score.} 
\label{fig:alpha}
\end{figure}

Figure~\ref{fig:alpha} illustrates the impact of $\alpha$ on retrieval performance. While the optimal $\alpha$ can vary across models, \textit{multilingual-e5-large} exhibits an inverted-U-shaped curve with a peak at $\alpha=0.6$.
In contrast, all other models show a monotonic increase, reaching their maximum performance at $\alpha=1.0$.
Crucially, \textsc{Shift} consistently outperforms the baseline across the entire range of $\alpha$.
This demonstrates that our method effectively mitigates language bias without hinging on a narrowly tuned hyperparameter, although selecting a model-specific $\alpha$ can further maximize the performance gains.

\begin{figure*}[t!]
\centering
\includegraphics[width=1.0\textwidth]
{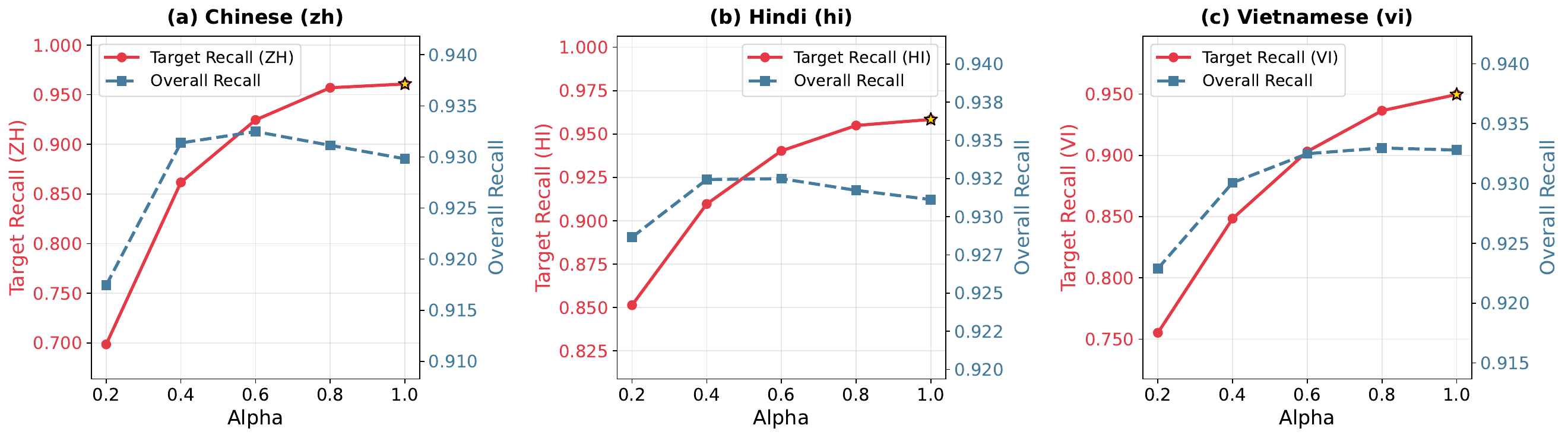}
\caption{Sensitivity analysis of the scale factor $\alpha$ for specific target languages~(Chinese, Hindi, Vietnamese). The solid red line represents performance on the specific target language (Target Recall@20), while the dashed blue line represents the overall system performance (Overall Recall@20). 
}
\label{fig:sensitivity}
\end{figure*}

\subsection{Sensitivity Analysis on \textsc{Shift} Magnitude}
\label{sec:analysis_alpha_sensitivity}

While Section~\ref{sec:analysis_alpha} explores the global impact of $\alpha$ across all languages, this section investigates its local impact on individual target languages and the resulting trade-off with overall system stability.
We conduct a fine-grained sensitivity analysis by varying the $\alpha$ value for a specific target language, while fixing the $\alpha$ values for other languages to their global optimal settings identified previously.

\paragraph{Evaluation Setup}
To analyze target-language sensitivity to the scale factor $\alpha$, we conduct experiments on two MLIR benchmarks, Belebele and XQuAD, using \textit{multilingual-e5-large} with English as the fixed source (query) language.
Based on Figure~\ref{fig:alpha}, we set the global optimal scale for \textit{multilingual-e5-large} to $\alpha=0.6$.
We experiment on three target languages (zh, hi, vi) and perform a per-language ablation: we vary $\alpha$ for the target language while fixing $\alpha=0.6$ for all other languages.
We report Overall Recall@20, computed over all relevant documents regardless of language, together with Target Recall@20 for each target language.
Specifically, unlike TLR@20 (which aggregates over all non-query languages), Target Recall@20 recomputes recall by restricting the ground-truth set to relevant documents in a single target language and treating relevant documents in other languages as non-relevant.
All metrics are averaged over the two benchmarks to assess generalization.

\paragraph{Results}
Figure~\ref{fig:sensitivity} illustrates the impact of $\alpha$ on retrieval performance. Regardless of the target language, we observe two consistent trends.
First, Target Recall@20 monotonically improves as $\alpha$ increases. This confirms the high precision of the directionality of the relative language vectors estimated via mMARCO. Geometrically, if the relative language vector fails to capture the language's characteristics properly, increasing $\alpha$ would move document embeddings away from the query language space, causing a sharp performance drop. Therefore, the continuous rise in target language performance indicates that the relative language vectors extracted by our method form a correct trajectory from the target language space to the source language space.

Second, Overall Recall@20 exhibits an inverted-U trend, decreasing beyond a certain $\alpha$.
This suggests over-shifting: pushing a single target language too aggressively can expose documents of that language in high ranks, but it simultaneously pushes down relevant documents in other non-target languages from the top ranks.
In other words, optimizing $\alpha$ for one language does not necessarily optimize overall retrieval across languages. Additional results for \textit{multilingual-e5-large} on other datasets are reported in Appendix~\ref{sec:appen_additional_alpha_experiment}, and results for additional retrievers are presented in Appendix~\ref{sec:appen_additional_alpha_experiment_models}.

\paragraph{Controllability and Robustness} 
Furthermore, these results suggest that $\alpha$ can serve as a controllable parameter in real-world environments. As observed, increasing $\alpha$ can explicitly boost the exposure of documents in a specific target language, even if it slightly compromises the overall ranking balance. This offers a practical advantage
when dynamic adjustment of result composition is needed based on user language preferences or regional characteristics. In essence, \textsc{Shift} becomes an effective tool for managing the trade-off between semantic alignment and language preference.

\section{Conclusion}
In this work, we demonstrate that state-of-the-art multilingual dense retrieval models suffer from severe language bias, disproportionately favoring query-language documents. To address this, we propose \textsc{Shift}, a training-free, index-side transformation that aligns document embeddings by subtracting a relative language vector estimated from parallel data. Furthermore, we introduce Target-Languages Recall@k (TLR@k) to rigorously quantify this language bias. Extensive experiments across diverse benchmarks and multiple model architectures confirm that \textsc{Shift} significantly improves MLIR performance, successfully redirecting the focus from linguistic surface forms to underlying semantics.

\section*{Limitations}
Despite its effectiveness, our approach has certain limitations. 

First, the estimation of language vectors relies on the mMARCO dataset, which contains machine-translated texts. Consequently, the quality of the shift vector may be influenced by the translation quality of the parallel corpus. 

Second, \textsc{Shift} requires the target documents to be pre-labeled with their language to apply the correct vector subtraction during indexing. Future work will focus on relaxing this dependency on explicit language labels and exploring purely unsupervised methods for language vector estimation.

Finally, \textsc{Shift} requires re-estimating the relative language vectors whenever the source language or the embedding model changes. In practice, however, this cost is incurred only once per setting: once computed, the vectors can be cached and reused across indexing runs and datasets as long as the same model and source language are maintained.

\section*{Ethics Statement}
This study utilizes publicly available datasets and open-source libraries, specifically sentence-transformers, for all experiments. We adhered to the licenses and terms of use associated with these resources. regarding the use of AI tools, we utilized AI assistants (e.g., ChatGPT, Gemini) exclusively for grammatical error correction and polishing the text to enhance readability. The scientific ideas and contributions presented in this paper are entirely our own.

\section*{Acknowledgments}
This research was supported by Basic Science Research Program through the National Research Foundation of Korea(NRF) funded by the Ministry of Education(NRF-2021R1A6A1A03045425). 
This work was supported by Institute for Information \& communications Technology Promotion(IITP) grant funded by the Korea government(MSIT) 
(RS-2024-00398115, Research on the reliability and coherence of outcomes produced by Generative AI).
This research was supported by Culture, Sports and Tourism R\&D Program through the Korea Creative Content Agency grant funded by the Ministry of Culture, Sports and Tourism in 2026 (Project Name: Development of an AI Agent Integrating Korean Language Knowledge for Personalized Language Consultation Services, Project Number: RS-2026-25506607, Contribution Rate: 25\%). 
This work was supported by Institute of Information \& communications Technology Planning \& Evaluation (IITP) under the artificial intelligence star fellowship support program to nurture the best talents (IITP-2026-RS-2025-02304828) grant funded by the Korea government(MSIT).

\bibliography{custom}

\clearpage


\appendix


\section{Evaluation Dataset Details}
\label{sec:evaluation_dataset_details}
In this work, we leverage multilingual Question Answering~(QA) datasets with parallel constructions, repurposed as retrieval tasks, to systematically assess MLIR performance. Since these datasets are originally designed for QA, the corresponding passages serve as exact positive documents within the retrieval framework. Consequently, datasets developed for QA are widely adopted for the evaluation of retrieval models in current literature~\cite{enevoldsen2025mmtebmassivemultilingualtext,lee2025geminiembeddinggeneralizableembeddings,qwen3embedding}.

\paragraph{Belebele} Belebele is a multilingual machine reading comprehension dataset spanning 122 language variants. Based on the FLORES-200 benchmark, each instance consists of a short passage and a corresponding multiple-choice question. For the retrieval task, we utilize the question stem as the query and the passage as the positive document. A defining feature of Belebele is its full parallelism: the entire set of questions and passages represents strictly parallel content across all 122 languages.
For our evaluation, we select the 14 languages that overlap with the mMARCO dataset: Arabic (ar), German (de), Dutch (nl), English (en), Spanish (es), French (fr), Hindi (hi), Indonesian (id), Italian (it), Japanese (ja), Portuguese (pt), Russian (ru), Vietnamese (vi), and Chinese (zh).

\paragraph{XQuAD} XQuAD is a cross-lingual QA benchmark consisting of a subset of the SQuAD v1.1 development set. The dataset was constructed by professionally translating both the questions and the paragraphs (contexts) from English into 10 target languages. This rigorous human translation process guarantees that XQuAD is fully parallel; every question (query) and paragraph (document) has an exact, semantically aligned counterpart in all other languages. 
We utilize the 8 languages intersecting with mMARCO for evaluation: Arabic (ar), German (de), English (en), Spanish (es), Hindi (hi), Russian (ru), Vietnamese (vi), and Chinese (zh).

\paragraph{MLQA} MLQA is a multi-way aligned extractive QA benchmark covering seven languages. Unlike Belebele or XQuAD, MLQA is designed to be multi-way parallel; each QA instance is aligned across a subset of languages (typically four) rather than the entire set. This structure results from mining parallel sentences where overlaps exist across varying language combinations to maximize linguistic diversity. Consequently, queries are also not fully parallel across all seven languages for every instance. Instead, for a given document, queries exist only in the specific subset of languages aligned with that document. Despite this partial parallelism, the dataset ensures that within each 4-way subset, the questions and contexts are semantically equivalent.
Since all 7 languages overlap with mMARCO, we utilize the complete language set for evaluation: Arabic (ar), German (de), English (en), Spanish (es), Hindi (hi), Vietnamese (vi), and Simplified Chinese (zh).

\paragraph{MultiEup-v2} Following \citet{bias1}, we utilize MultiEup-v2 as an MLIR benchmark. The task is defined as retrieving the relevant parliamentary speech segment (document) given a subject descriptor (query). The dataset is constructed from European Parliament proceedings, where professional archivists manually assign official subject descriptors (e.g., `International Human Rights') to specific speech segments. Since the descriptors are professionally translated into 24 languages, the same information can be expressed as multilingual queries. In this framework, the ground truth is the specific speech text that was originally tagged with the descriptor by human experts.
We use the 6 languages that overlap with mMARCO: German (de), English (en), Spanish (es), French (fr), Italian (it), and Portuguese (pt).

We note that Belebele, XQuAD, and MLQA are utilized in MMTEB~\cite{enevoldsen2025mmtebmassivemultilingualtext}, while MultiEup-v2 is adopted to evaluate MLIR tasks following~\citet{bias1}. 
We do not include Mr. TyDi~\citep{mrtidy} or MIRACL~\citep{miracl} in our evaluation, since their relevance annotations are not explicitly labeled across languages and the datasets are neither parallel nor aligned at the document level, making them unsuitable for measuring retrieval performance under our MLIR setup.

\begin{table*}[ht!]
\centering
\resizebox{\textwidth}{!}{
\begin{tabular}{l|l|l|c}
\toprule
\textbf{Model} & \textbf{query prefix} & \textbf{document prefix} & \textbf{\# of trained\_langs} \\
\midrule
embeddinggemma-300m & task: search result | query: \{query\} & title: none | text: \{document\} & 100+ \\
\midrule[0.2pt]
multilingual-e5-large & query: \{query\} & passage: \{document\} & 94 \\
\midrule[0.2pt]
bge-m3 & - & - & 100+ \\
\midrule[0.2pt]
Qwen3-Embedding-0.6B & \begin{tabular}{@{}l@{}}Instruct: Given a web search query, retrieve relevant \\ passages that answer the query \textbackslash nQuery: \{query\}\end{tabular} & - & 100+ \\
\midrule[0.2pt]
llama-nemotron-embed-1b-v2 & query: \{query\} & passage: \{document\} & 26 \\
\midrule[0.2pt]
gte-Qwen2-1.5B-instruct & \begin{tabular}{@{}l@{}}Instruct: Given a web search query, retrieve relevant \\ passages that answer the query \textbackslash nQuery: \{query\}\end{tabular} & - & 33 \\
\bottomrule
\end{tabular}
}
\caption{Comparison of Multilingual Embedding Models}
\label{tab:model_details}
\end{table*}

\section{Implementation Details}
\label{sec:implementation_details}
\paragraph{Model Details} We utilize the \texttt{sentence-transformers}\footnote{\url{https://github.com/huggingface/sentence-transformers}} library for inference. 
For \textit{bge-m3}, we exclusively employ the dense embedding component. 
The maximum sequence length is set to 512 tokens for all models. 
To ensure optimal performance, we strictly adhere to the specific prefixes and instruction prompts required by each retrieval model. Language vectors are computed from unnormalized embeddings, and during evaluation, we apply language shifts to the unnormalized embeddings before computing cosine similarity which internally normalizes the vectors. We provide the details in Table~\ref{tab:model_details}.

\paragraph{Hardware Details} We conducted our experiments using an Intel Xeon Gold 6230R @2.10GHz CPU, 376GB RAM, and 8 NVIDIA RTX A6000 48GB GPUs. The software environment included nvidia driver, CUDA, and PyTorch, running on Ubuntu 20.04.6 LTS.

\section{Comparison with Baselines}
\label{sec:baseline}

To situate \textsc{Shift} among prior training-free approaches for mitigating language bias, we compare it with two representative post-hoc baselines on three retrievers: \textit{embeddinggemma-300m}, \textit{multilingual-e5-large}, and \textit{bge-m3}. Table~\ref{tab:appen_retrieval_performance_baseline} reports nDCG@20 and TLR@20 on four MLIR benchmarks.

\subsection{Baselines}

\paragraph{Language-wise centering (Centering)}
\citet{libovicky-etal-2020-language} propose an unsupervised centering procedure motivated by the view that multilingual representations contain both language-specific and language-neutral components.
Let $\mathbf{x} \in \mathbb{R}^{d}$ denote the $d$-dimensional embedding produced by the retriever for an input text (a query or a document).
Given a language $\ell$, the method estimates a language centroid as the mean embedding of texts in the language,
\begin{equation*}
\boldsymbol{\mu}_{\ell} = \frac{1}{N_{\ell}} \sum_{i=1}^{N_{\ell}} \mathbf{x}^{(\ell)}_i
\end{equation*}
and applies language-wise re-centering by subtracting the centroid from each embedding:
\begin{equation*}
\tilde{\mathbf{x}} = \mathbf{x} - \boldsymbol{\mu}_{\ell}
\end{equation*}
We use this centering variant as a training-free baseline and apply the same transformation to both queries and documents based on their language.

\paragraph{Language Information Removal (LIR)}
\citet{langdetect1} introduce Language Information Removal (LIR), a post-training method that suppresses language identification information by removing dominant language directions in the representation space.
For each language $\ell$, LIR constructs an embedding matrix $\mathbf{X}_{\ell} \in \mathbb{R}^{N_{\ell} \times d}$ by stacking embeddings of texts written in $\ell$.
It then computes the Singular Value Decomposition (SVD),
\begin{equation*}
\mathbf{X}_{\ell} = \mathbf{U}_{\ell}\mathbf{\Sigma}_{\ell}\mathbf{V}_{\ell}^{\top}
\end{equation*}
and takes the top-$r$ right-singular vectors $\mathbf{V}_{\ell,r} \in \mathbb{R}^{d \times r}$ as principal directions associated with language identity.
Each embedding is updated by subtracting its projection onto the subspace spanned by these directions:
\begin{equation*}
\tilde{\mathbf{x}} = \mathbf{x} - \mathbf{V}_{\ell,r}\mathbf{V}_{\ell,r}^{\top}\mathbf{x}
\end{equation*}
Equivalently, this removes components along the top directions, $\tilde{\mathbf{x}} = \mathbf{x} - \sum_{j=1}^{r} (\mathbf{v}_j^{\top}\mathbf{x})\mathbf{v}_j$, where $\mathbf{v}_j$ is the $j$-th column of $\mathbf{V}_{\ell,r}$.

\subsection{Implementation}
For a controlled comparison, we compute all language-dependent statistics required by the baselines from mMARCO using the same retriever under evaluation.

\paragraph{Centering}
For each model and each language $\ell$, we embed mMARCO texts written in $\ell$ and estimate the centroid $\boldsymbol{\mu}_{\ell}$ as the mean of their embeddings.
At indexing time, each document embedding is transformed as $\mathbf{x} \mapsto \mathbf{x}-\boldsymbol{\mu}_{\ell}$.
At retrieval time, we apply the same transformation to the query embedding before nearest-neighbor search.

\begin{table*}[h!]
\centering
\resizebox{1.0\textwidth}{!}{%
\begin{tabular}{l|cccccccc||cc}
\toprule
\multicolumn{1}{c|}{\multirow{2}{*}{\textbf{Model}}} & \multicolumn{2}{c}{\textbf{Belebele}} & \multicolumn{2}{c}{\textbf{MLQA}} & \multicolumn{2}{c}{\textbf{XQuAD}} & \multicolumn{2}{c||}{\textbf{MultiEup-v2}} & \multicolumn{2}{c}{\textbf{AVG}} \\ 
\cmidrule(lr){2-3} \cmidrule(lr){4-5} \cmidrule(lr){6-7} \cmidrule(lr){8-9} \cmidrule(lr){10-11} 
\multicolumn{1}{c|}{} & nDCG & TLR & nDCG & TLR & nDCG & TLR & nDCG & TLR & nDCG & TLR \\ \midrule

embeddinggemma-300m (base) & 0.924 & 0.922 & 0.664 & 0.706 & 0.947 & 0.966 & 0.495 & 0.318 & 0.758 & 0.728 \\
Centering~\citep{libovicky-etal-2020-language} & 0.924 & 0.924 & 0.669 & 0.727 & 0.947 & 0.967 & 0.508 & 0.363 & 0.762 & 0.745 \\
LIR~\citep{langdetect1} & \textbf{0.926} & 0.925 & \textbf{0.671} & 0.728 & \textbf{0.951} & 0.970 & 0.509 & 0.357 & 0.764 & 0.745 \\
\textbf{\textsc{Shift}} & \textbf{0.926} & \textbf{0.926} & \textbf{0.671} & \textbf{0.736} & \textbf{0.951} & \textbf{0.971} & \textbf{0.511} & \textbf{0.371} & \textbf{0.765} & \textbf{0.751} \\ \midrule

multilingual-e5-large (base) & 0.816 & 0.798 & 0.494 & 0.380 & 0.855 & 0.886 & 0.367 & 0.095 & 0.633 & 0.540 \\
Centering~\citep{libovicky-etal-2020-language} & 0.889 & 0.877 & 0.622 & 0.624 & 0.930 & 0.942 & 0.443 & 0.196 & 0.721 & 0.660 \\
LIR~\citep{langdetect1} & 0.891 & 0.880 & 0.623 & 0.625 & 0.931 & 0.943 & \textbf{0.444} & 0.261 & 0.722 & 0.677 \\
\textbf{\textsc{Shift}} & \textbf{0.913} & \textbf{0.898} & \textbf{0.649} & \textbf{0.654} & \textbf{0.944} & \textbf{0.956} & 0.442 & \textbf{0.268} & \textbf{0.737} & \textbf{0.694} \\ \midrule

bge-m3 (base) & 0.874 & 0.864 & 0.606 & 0.637 & 0.923 & 0.942 & 0.455 & 0.293 & 0.715 & 0.684 \\
Centering~\citep{libovicky-etal-2020-language} & 0.849 & 0.838 & 0.598 & 0.626 & 0.912 & 0.931 & 0.454 & 0.295 & 0.703 & 0.673 \\
LIR~\citep{langdetect1} & 0.862 & 0.853 & 0.605 & 0.637 & 0.920 & 0.939 & 0.458 & \textbf{0.300} & 0.711 & 0.682 \\
\textbf{\textsc{Shift}} & \textbf{0.876} & \textbf{0.868} & \textbf{0.609} & \textbf{0.643} & \textbf{0.925} & \textbf{0.945} & \textbf{0.460} & \textbf{0.300} & \textbf{0.717} & \textbf{0.689} \\ \bottomrule
\end{tabular}
}
\caption{Top-20 retrieval performance (nDCG@20, TLR@20) of base models, \citet{libovicky-etal-2020-language}, \citet{langdetect1}, and \textsc{Shift} across four MLIR benchmarks, evaluated with English as the source query language.}
\label{tab:appen_retrieval_performance_baseline}
\end{table*}

\paragraph{LIR}
For each model and each language $\ell$, we construct the embedding matrix $\mathbf{X}_{\ell}$ by stacking embeddings of mMARCO texts in $\ell$, and compute its SVD to obtain $\mathbf{V}_{\ell,r}$.
Following \citet{langdetect1}, which reports that removing only the top component performs best in their setting, we set $r=1$ throughout.
At indexing time, each document embedding is transformed as $\mathbf{x} \mapsto \mathbf{x}-\mathbf{V}_{\ell,r}\mathbf{V}_{\ell,r}^{\top}\mathbf{x}$, and we apply the same transformation to the query embedding at retrieval time.

\subsection{Results \& Analysis}

\paragraph{Results} Table~\ref{tab:appen_retrieval_performance_baseline} shows that \textsc{Shift} yields the strongest average performance across the three retrievers when considering both nDCG@20 and TLR@20.
The difference is most pronounced for \textit{multilingual-e5-large}, where the base model exhibits low TLR (AVG TLR@20 of 0.540).
Both centering and LIR increase TLR@20 substantially, while \textsc{Shift} achieves the highest average TLR@20 (0.694) and also improves average nDCG@20.
For \textit{embeddinggemma-300m}, all methods yield modest but consistent improvements, with \textsc{Shift} attaining the best averages.
For \textit{bge-m3}, the improvements are smaller and both baselines are slightly worse than the base model on average, suggesting that removing global language components is not uniformly beneficial when the embedding space is already relatively well aligned.
This behavior is consistent with prior observations that removing dominant components may also remove semantic information~\citep{huang2023soft}.
Overall, the comparison indicates that \textsc{Shift} provides consistent gains across models, while maintaining strong target-language exposure as measured by TLR@20.

\paragraph{Inference-time practicality}
While all compared methods are training-free, the baselines require additional query-time transformations: centering subtracts a language-dependent centroid, and LIR removes language identification components via an additional projection step.
In contrast, \textsc{Shift} can be applied to document embeddings offline during indexing, and the query-side operation is fixed given a source language.
As a result, \textsc{Shift} reduces inference-time overhead and is simpler to deploy in latency-sensitive retrieval systems.

\begin{figure*}[h!]
    \centering
    \includegraphics[width=1.0\linewidth]
    {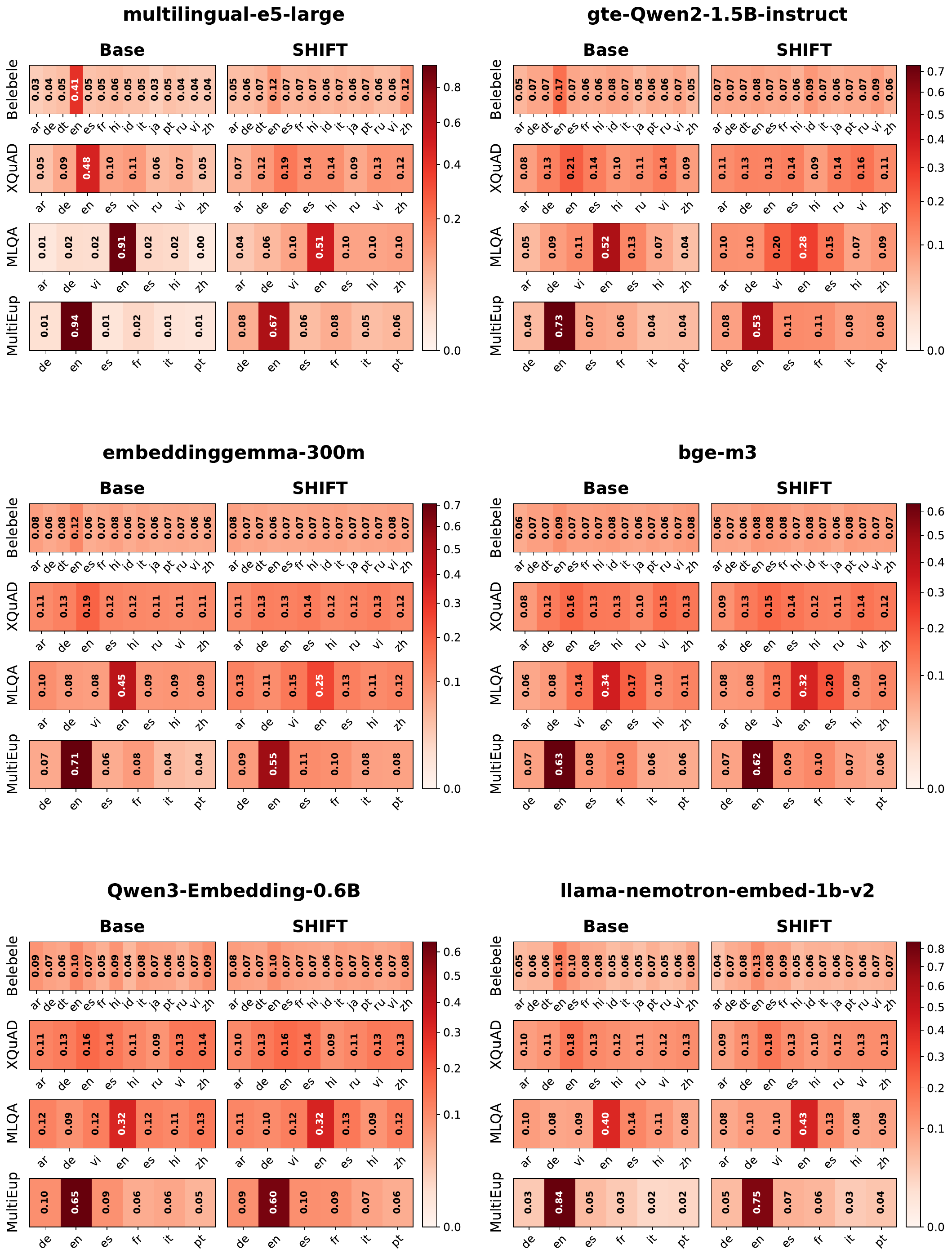}
    \caption{Extended visualization of language distribution heatmaps for all evaluated models.} 
    \label{fig:appen_heatmap}
\end{figure*}

\section{Additional Language Distribution Analysis}
\label{sec:appen_heatmap}

Figure~\ref{fig:appen_heatmap} provides a comprehensive visualization of language distribution heatmaps for all evaluated models across the four benchmarks.
Consistent with the findings in Section~\ref{sec:analysis_distribution}, the baseline results (left panels) universally exhibit a strong English-centric bias, regardless of the model architecture.
Whether using encoder-based models like \textit{embeddinggemma-300m} or decoder-based models like \textit{gte-Qwen2-1.5B-instruct}, the retrieval density is disproportionately concentrated in the 'en' column, leaving other target languages marginalized.
However, applying \textsc{Shift} (right panels) alleviates this skew across all architectures.
The density, previously locked on English, effectively disperses to the relevant target languages, resulting in a more semantically balanced distribution.
This visual evidence further corroborates the quantitative improvements reported in Table~\ref{tab:tlr}, confirming that \textsc{Shift} serves as a robust and universal solution for rectifying language bias, independent of the underlying model type.

\begin{table*}[h]
\centering
\normalsize
\resizebox{1.0\textwidth}{!}
{
\begin{tabular}{l|ccc|ccc|ccc}
\toprule
\textbf{Source Language} & \multicolumn{3}{c|}{\textbf{ZH}} & \multicolumn{3}{c|}{\textbf{VI}} & \multicolumn{3}{c}{\textbf{HI}} \\ 
\midrule
\textbf{Datasets} & \textbf{Belebele} & \textbf{MLQA} & \textbf{XQuAD} & \textbf{Belebele} & \textbf{MLQA} & \textbf{XQuAD} & \textbf{Belebele} & \textbf{MLQA} & \textbf{XQuAD} \\ 
\midrule
\# of Languages & 14 & 7 & 8 & 14 & 7 & 8 & 14 & 7 & 8 \\
Fully Parallel & O & X & O & O & X & O & O & X & O \\
\midrule
\# of Queries & 900 & 5,136 & 1,190 & 900 & 5,495 & 1,190 & 900 & 4,916 & 1,190 \\
\# of Documents & 6,832 & 17,165 & 1,920 & 6,832 & 18,150 & 1,920 & 6,832 & 15,639 & 1,920 \\
\midrule
\# Avg. $d^+/q$ & 14 & 3.79 & 8 & 14 & 3.75 & 8 & 14 & 3.77 & 8 \\
\bottomrule
\end{tabular}
}
\caption{Detailed statistics of the datasets used for the multi-source language evaluation (Section~\ref{sec:results_otherlangs}), where Chinese (ZH), Vietnamese (VI), and Hindi (HI) serve as the source query languages.}
\label{tab:dataset_stats_multisrc}
\end{table*}

\section{Multi-Source Language Evaluation Dataset}
\label{sec:multi_source_language_evaluation_dataset}
Table~\ref{tab:dataset_stats_multisrc} summarizes the detailed statistics of the datasets configured for the non-English source query evaluation, where Chinese (ZH), Vietnamese (VI), and Hindi (HI) serve as the source query languages.
As shown in the table, fully parallel datasets such as Belebele and XQuAD maintain consistent statistics across all source languages, including English in Table~\ref{tab:dataset_stats}.
In contrast, MLQA exhibits variations in the number of queries and documents depending on the source language.
This is due to its partially parallel nature, where the availability of aligned query-passage pairs differs across language subsets.

\section{Additional Evaluation on NeuCLIR2023}
\label{sec:appen_neuclir}

To further evaluate \textsc{Shift} beyond parallel or translation-derived benchmarks, we conduct an additional experiment on a NeuCLIRBench dataset~\citep{lawrie2025neuclirbenchmodernevaluationcollection}, specifically NeuCLIR2023RetrievalHardNegatives~\footnote{\url{https://huggingface.co/datasets/mteb/NeuCLIR2023RetrievalHardNegatives}}.
Unlike our main experiments where relevance is constructed from semantically aligned multilingual passages, this benchmark represents a more open multilingual retrieval setting. The original task uses English queries and retrieves from a mixed multilingual document pool containing Chinese, Russian, and Persian documents. Since \textsc{Shift} requires relative language vectors estimated for each target language, we evaluate the subset of Chinese and Russian documents, which overlap with the languages supported by our vector-estimation corpus. We report nDCG@20 and apply the same model-specific scale factor $\alpha$ selected in the main experiments, without any dataset-specific tuning.

\begin{table}[h]
\centering
\resizebox{\columnwidth}{!}{
\begin{tabular}{lrrr}
\toprule
\textbf{Model} & \textbf{Base} & \textbf{\textsc{Shift}} & \textbf{$\Delta$} \\
\midrule
embeddinggemma-300m & 0.514 & \textbf{0.520} & +0.006 \\
multilingual-e5-large & 0.409 & \textbf{0.448} & +0.039 \\
bge-m3 & \textbf{0.472} & 0.470 & -0.002 \\
Qwen3-Embedding-0.6B & 0.461 & \textbf{0.467} & +0.006 \\
llama-nemotron-embed-1b-v2 & \textbf{0.515} & \textbf{0.515} & +0.000 \\
gte-Qwen2-1.5B-instruct & 0.468 & \textbf{0.483} & +0.015 \\
\midrule
Average & 0.4731 & \textbf{0.4839} & +0.0108 \\
\bottomrule
\end{tabular}
}
\caption{Retrieval performance (nDCG@20) on the NeuCLIR2023 benchmark.}
\label{tab:appen_neuclir}
\end{table}

As shown in Table~\ref{tab:appen_neuclir}, \textsc{Shift} improves the average nDCG@20 from 0.4731 to 0.4839, yielding gains for 5 out of 6 evaluated retrievers. The most substantial improvement is observed for \textit{multilingual-e5-large} (+0.039), which corresponds to the model exhibiting strong language bias in our previous analyses. Conversely, the performance of \textit{bge-m3} changes only marginally (-0.002). This aligns with our observation that \textit{bge-m3} already possesses relatively strong multilingual alignment, leaving less language-induced offset for \textsc{Shift} to correct.

\section{Robustness to Imperfect Language Identification}
\label{sec:appen_robustness}

Because \textsc{Shift} applies language-specific transformations during the indexing stage, its practical deployment depends on identifying the document languages. In our primary experiments, we utilized ground-truth document language labels to strictly isolate the effect of the proposed transformation. To evaluate a more realistic deployment scenario, we conduct a robustness experiment where document languages are automatically predicted using an off-the-shelf FastText language identification model\footnote{\url{https://huggingface.co/facebook/fasttext-language-identification}}, and \textsc{Shift} is applied based on these predicted labels.

\begin{figure*}[h!]
    \centering
    \includegraphics[width=0.75\textwidth]
    {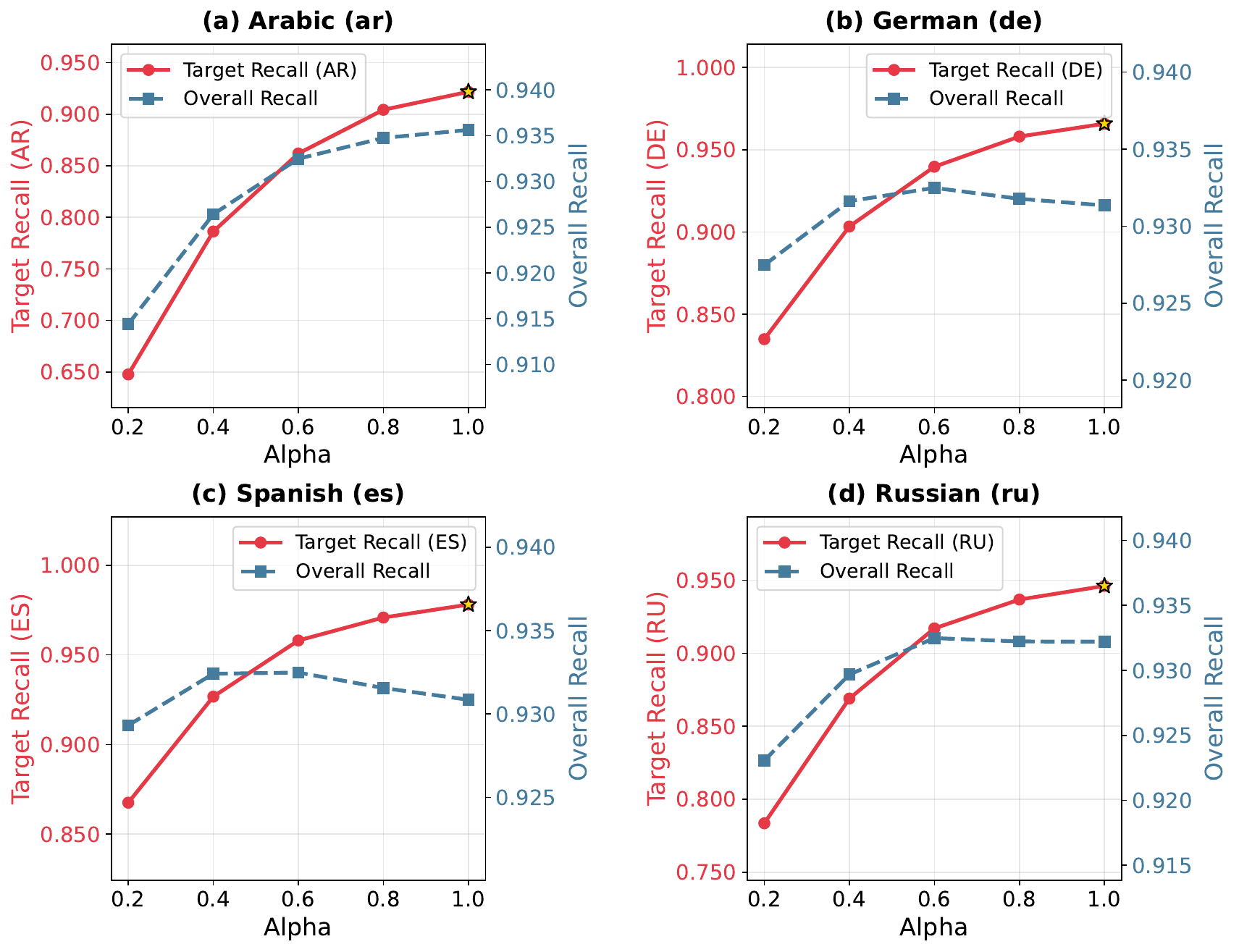}
    \caption{Extended sensitivity analysis of the scale factor $\alpha$ across remaining 4 target languages.} 
    \label{fig:appen_e5_alpha_sensitivity}
\end{figure*}

We evaluate \textit{multilingual-e5-large} using the same scale factor ($\alpha=0.6$) as in the main experiments. We compare three settings: \textbf{Base}, \textbf{GT-\textsc{Shift}} (using ground-truth document language labels), and \textbf{FT-\textsc{Shift}} (using FastText-predicted document language labels). The evaluation spans the four MLIR benchmarks from the main experiments, as well as the NeuCLIR 2023 setting. We report Language Identification (LID) accuracy, defined as the percentage of corpus documents whose FastText-predicted language matches the ground-truth language.

\begin{table}[h]
\centering
\resizebox{\columnwidth}{!}{
\begin{tabular}{lrrrrr}
\toprule
\textbf{Benchmark} & \textbf{Base} & \textbf{GT-\textsc{Shift}} & \textbf{FT-\textsc{Shift}} & \textbf{$\Delta$ (FT$-$GT)} & \textbf{LID Acc.} \\
\midrule
Belebele & 0.623 & 0.813 & 0.807 & -0.006 & 99.4\% \\
MLQA & 0.436 & 0.548 & 0.534 & -0.016 & 93.4\% \\
MultiEup & 0.309 & 0.389 & 0.389 & +0.000 & 99.1\% \\
XQuAD & 0.717 & 0.864 & 0.846 & -0.018 & 96.3\% \\
NeuCLIR 2023 & 0.409 & 0.448 & 0.421 & -0.027 & 50.2\% \\
\midrule
Average & 0.499 & 0.613 & 0.600 & -0.013 & 87.7\% \\
\bottomrule
\end{tabular}
}
\caption{Robustness of \textsc{Shift} to imperfect language identification using FastText on \textit{multilingual-e5-large}. Performance is measured by nDCG@20.}
\label{tab:appen_robustness}
\end{table}

The results are summarized in Table~\ref{tab:appen_robustness}. \textbf{FT-\textsc{Shift}} consistently outperforms the \textbf{Base} model across all benchmarks, demonstrating that \textsc{Shift} remains effective even when relying on automatically predicted language labels. On average, \textbf{FT-\textsc{Shift}} improves nDCG@20 from 0.499 to 0.600, maintaining performance very close to \textbf{GT-\textsc{Shift}} with only a minimal average drop of 0.013. 

The most noticeable degradation occurs on NeuCLIR 2023, where the FastText LID accuracy is only 50.2\% (primarily due to a 7.4\% accuracy drop on Chinese documents in this specific corpus). Nevertheless, even in this challenging setting, \textbf{FT-\textsc{Shift}} still surpasses the \textbf{Base} model (0.421 vs. 0.409). In contrast, on MultiEup, where all document languages are accurately identified, \textbf{FT-\textsc{Shift}} is virtually identical to \textbf{GT-\textsc{Shift}}. Ultimately, in deployment, \textsc{Shift} only requires predicted document languages at indexing time. Because both the language prediction and the corresponding vector transformation are applied offline, they introduce no additional overhead.

\begin{figure*}[h!]
    \centering
    \includegraphics[width=0.9\textwidth]{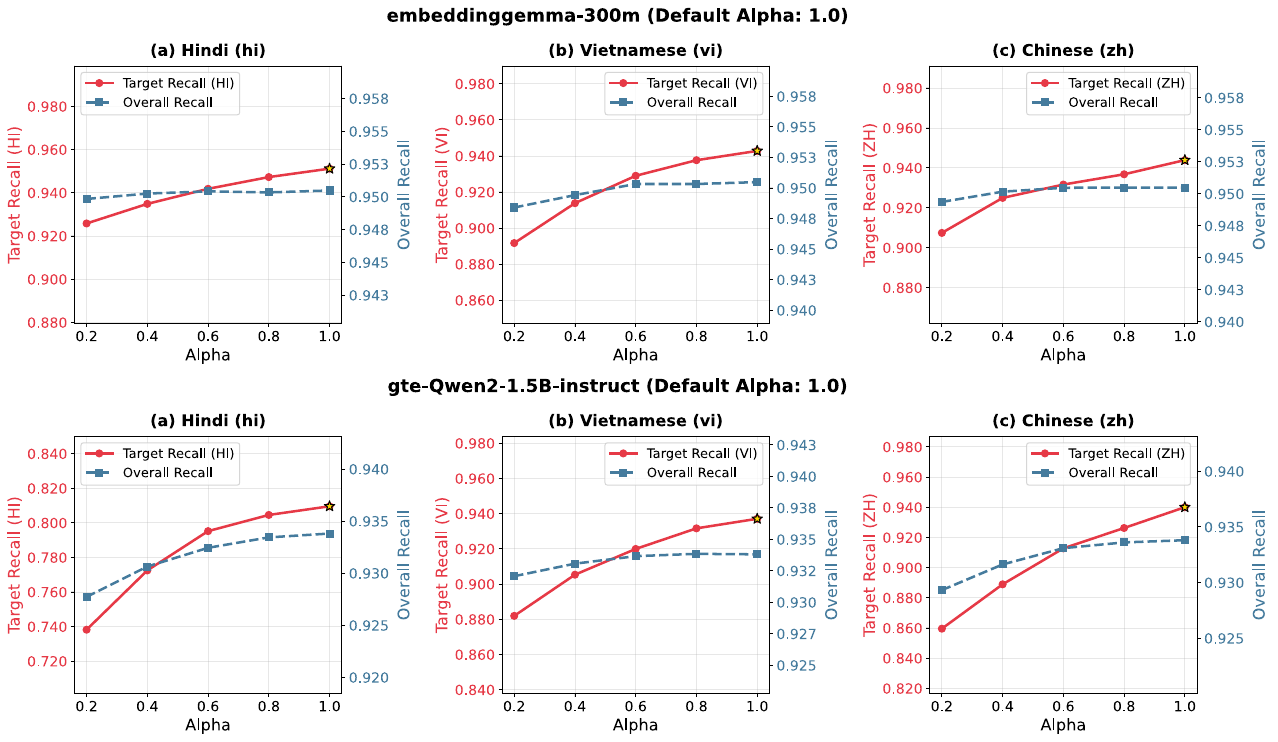}
    \caption{Sensitivity analysis with an Aggressive Default ($\alpha_{default}=1.0$). We vary the $\alpha$ for the target language while fixing $\alpha=1.0$ for all other languages.} 
    \label{fig:additional_alpha_sensitivity_1.0}
\end{figure*}

\section{Additional Sensitivity Analysis for \textit{multilingual-e5-large}}
\label{sec:appen_additional_alpha_experiment}

We extend the sensitivity analysis presented in Section~\ref{sec:analysis_alpha_sensitivity} to the remaining four languages available in the intersection of mMARCO, Belebele, and XQuAD: Arabic (ar), German (de), Spanish (es), and Russian (ru).
The results are visualized in Figure~\ref{fig:appen_e5_alpha_sensitivity}.

As illustrated, the performance trajectories exhibit identical patterns to those reported in the main experiments.
First, the Target Recall@20 (red line) monotonically increases as $\alpha$ grows, reaffirming the precise directionality of the estimated relative language vectors across diverse language families.
Second, the Overall Recall@20 (blue dashed line) displays a characteristic inverted-U or saturation trend.
This consistent behavior across all evaluated languages further validates that while increasing $\alpha$ effectively boosts target-specific retrieval, an optimal threshold exists to maintain the global stability of the multilingual ranking system.

\section{Sensitivity Analysis on Diverse Architectures and Base Settings}
\label{sec:appen_additional_alpha_experiment_models}

To verify the universality of our findings beyond a single model architecture, we extend the sensitivity analysis to two distinct models: \textit{embeddinggemma-300m} (encoder-based) and \textit{gte-Qwen2-1.5B-instruct} (decoder-based). Furthermore, to investigate how the global shift intensity influences the local sensitivity of a specific target language, we conduct experiments under two extreme baseline settings:
\begin{itemize}
    \item Aggressive Default ($\alpha_{default}=1.0$): All non-target languages are shifted maximally. This simulates a fully transformed, source-aligned embedding space.
    \item Conservative Default ($\alpha_{default}=0.1$): All non-target languages are shifted minimally. This simulates a scenario where the global embedding space remains close to the original, biased distribution.
\end{itemize}

In both settings, we sweep the $\alpha$ for the specific target language (Chinese, Hindi, Vietnamese) from 0.2 to 1.0 and observe the trade-off between Target Recall and Overall Recall.

\paragraph{Aggressive Default}
Figure~\ref{fig:additional_alpha_sensitivity_1.0} presents the results when the base system is fully aligned to the source language.
In this setting, the continuous improvement in Target Recall and the high stability of Overall Recall are expected outcomes.
As demonstrated in Section~\ref{sec:analysis_alpha_sensitivity}, these models achieve their global optimum at $\alpha=1.0$.
Therefore, increasing the specific target $\alpha$ towards 1.0 effectively moves the target language embeddings closer to the source language's semantic space, reinforcing the semantic signals that were previously diluted by language barriers.
This confirms that the performance gains are a natural consequence of the embeddings converging to their optimal geometric alignment, where both the target-specific and global retrieval capabilities are maximized.

\begin{figure*}[h!]
    \centering
    \includegraphics[width=0.9\textwidth]{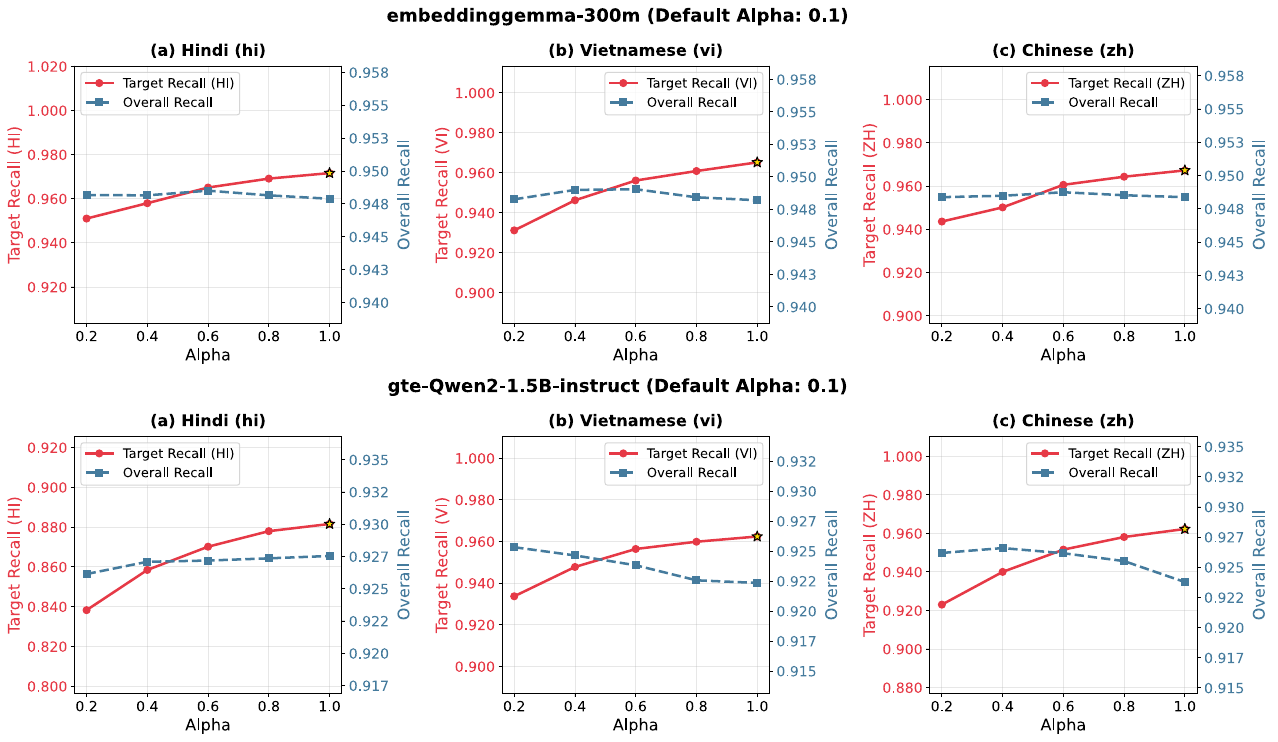}
    \caption{Sensitivity analysis with a Conservative Default ($\alpha_{default}=0.1$). We vary the $\alpha$ for the target language while fixing $\alpha=0.1$ for all other languages.} 
    \label{fig:additional_alpha_sensitivity_0.1}
\end{figure*}

\paragraph{Conservative Default}
Figure~\ref{fig:additional_alpha_sensitivity_0.1} illustrates the results when the base shift is minimal.
In this setting, the base system retains a strong bias towards the source language, typically suppressing target language documents.
However, as we increase the $\alpha$ for the specific target language, we observe a dramatic surge in Target Recall (solid red line).
This result is particularly significant from the perspective of controllability discussed in Section~\ref{sec:analysis_alpha_sensitivity}.
It demonstrates that even if the global system remains conservative, we can effectively "boost" the exposure of a specific language by selectively increasing its shift magnitude.
Although this targeted boosting involves a trade-off with Overall Recall (dashed blue line), it empirically confirms that \textsc{Shift} provides a powerful mechanism to forcefully prioritize specific languages according to user requirements.


\section{M2MIR: Multilingual-to-Multilingual Information Retrieval}
\label{sec:m2mir}

While standard MLIR tasks typically assume a monolingual query retrieving from a multilingual collection (one-to-many), real-world search scenarios often involve users issuing queries in diverse languages to find information across a globally mixed corpus.
To evaluate \textsc{Shift} in this more practical and challenging environment, we introduce the M2MIR (Multilingual-to-Multilingual Information Retrieval) scenario, where both queries and documents originate from multiple languages.

\subsection{Experimental Setup}
\begin{table}[ht!]
\centering
\normalsize
\resizebox{1.0\linewidth}{!}
{
\begin{tabular}{l|cccc}
\toprule
\textbf{Data Statistics} & \textbf{Belebele} & \textbf{MLQA} & \textbf{XQuAD} & \textbf{MultiEup-v2} \\
\midrule
\# of Languages in queries & \makecell{14} & \makecell{7} & \makecell{8} & \makecell{6} \\
\# of Languages in collection & \makecell{14} & \makecell{7} & \makecell{8} & \makecell{6} \\
Fully Parallel & O & X & O & X \\
\midrule
\# of Queries & 12,600 & 42,245 & 9,520 & 10,686 \\
\# of Documents & 6,832 & 36,799 & 1,920 & 35,444 \\
\midrule
\# of $d^+/q$ & 14 & 3.65 & 8 & 19.9 \\
\# of $d^+_{en}/q$ & 1 & 1 & 1 & 7.91 \\
\bottomrule
\end{tabular}
}
\caption{Statistics of the datasets configured for the M2MIR evaluation.}
\label{tab:dataset_stat_mmlir}
\end{table}
We reconfigure the test sets of Belebele, MLQA, XQuAD, and MultiEuP-v2 to include queries from all available languages overlapping with mMARCO, rather than restricting them to a single source language.
The detailed statistics of this constructed dataset are summarized in Table~\ref{tab:dataset_stat_mmlir}.
In this setting, the retrieval pool contains documents in varying languages, while the query stream simultaneously comprises a diverse mixture of languages.
For all \textsc{Shift} operations in this scenario, we use English as the anchor (reference) language, i.e., we align embeddings toward the English embedding space.
We set the scale factor $\alpha$ to 1.0 to apply the full shift toward the anchor space.
Consistent with the main experiments, we employ nDCG@20 and Recall@20 as the primary evaluation metrics.
We compare two strategies for applying \textsc{Shift} to align the embedding space toward English:

\begin{itemize}
    \item \textit{Doc Only}: We apply \textsc{Shift} exclusively to document embeddings. Specifically, we subtract the relative language vector corresponding to each document's language, aligning it with the English space. The query embeddings remain unmodified in their original language spaces.
    \item \textit{Query and Doc}: We apply \textsc{Shift} to both queries and documents. Queries in non-English languages are also shifted towards the English space using their respective relative language vectors. This ensures that both the queries and the documents reside in the unified English-centric geometry.
\end{itemize}

\begin{table*}[ht!]
\centering
\resizebox{\textwidth}{!}{
\begin{tabular}{l|cc|cc|cc|cc||cc}
\toprule
\multicolumn{1}{c|}{\multirow{2}{*}{Model}} & \multicolumn{2}{c|}{Belebele} & \multicolumn{2}{c|}{MLQA} & \multicolumn{2}{c|}{XQuAD} & \multicolumn{2}{c||}{MultiEup-v2} & \multicolumn{2}{c}{AVG} \\ \cmidrule(lr){2-3} \cmidrule(lr){4-5} \cmidrule(lr){6-7} \cmidrule(lr){8-9} \cmidrule(lr){10-11}
\multicolumn{1}{c|}{} & nDCG & Recall & nDCG & Recall & nDCG & Recall & nDCG & Recall & nDCG & Recall \\ \midrule
embeddinggemma-300m & 0.875 & 0.878 & 0.622 & 0.696 & 0.913 & 0.948 & 0.457 & 0.402 & 0.717 & 0.731 \\
\quad + \textsc{Shift} (\textit{Doc Only}) & 0.892 & 0.895 & 0.630 & 0.707 & 0.928 & 0.957 & 0.490 & 0.428 & 0.735 & 0.747 \\
\quad + \textsc{Shift} (\textit{Query and Doc}) & \textbf{0.893} & \textbf{0.896} & \textbf{0.630} & \textbf{0.707} & \textbf{0.928} & \textbf{0.957} & \textbf{0.494} & \textbf{0.431} & \textbf{0.736} & \textbf{0.748} \\ \midrule
multilingual-e5-large & 0.623 & 0.601 & 0.436 & 0.452 & 0.717 & 0.771 & 0.309 & 0.256 & 0.521 & 0.520 \\
\quad + \textsc{Shift} (\textit{Doc Only}) & 0.860 & 0.856 & 0.594 & 0.664 & 0.914 & 0.947 & \textbf{0.414} & \textbf{0.355} & 0.695 & \textbf{0.706} \\
\quad + \textsc{Shift} (\textit{Query and Doc}) & \textbf{0.862} & \textbf{0.860} & \textbf{0.597} & \textbf{0.668} & \textbf{0.916} & \textbf{0.948} & 0.409 & 0.347 & \textbf{0.696} & \textbf{0.706} \\ \midrule
llama-nemotron-embed-1b-v2 & 0.541 & 0.535 & 0.449 & 0.519 & 0.851 & 0.899 & 0.218 & 0.186 & 0.515 & 0.535 \\
\quad + \textsc{Shift} (\textit{Doc Only}) & 0.604 & 0.606 & 0.465 & 0.541 & 0.874 & \textbf{0.918} & 0.258 & 0.226 & 0.550 & 0.573 \\
\quad + \textsc{Shift} (\textit{Query and Doc}) & \textbf{0.617} & \textbf{0.618} & \textbf{0.468} & \textbf{0.543} & \textbf{0.876} & \textbf{0.918} & \textbf{0.268} & \textbf{0.235} & \textbf{0.557} & \textbf{0.579} \\ \midrule
gte-Qwen2-1.5B-instruct & 0.793 & 0.790 & 0.518 & 0.577 & 0.828 & 0.875 & 0.442 & 0.388 & 0.645 & 0.657 \\
\quad + \textsc{Shift} (\textit{Doc Only}) & 0.846 & 0.845 & 0.558 & 0.633 & 0.886 & 0.923 & \textbf{0.478} & \textbf{0.419} & 0.692 & 0.705 \\
\quad + \textsc{Shift} (\textit{Query and Doc}) & \textbf{0.853} & \textbf{0.852} & \textbf{0.566} & \textbf{0.642} & \textbf{0.892} & \textbf{0.927} & 0.477 & 0.417 & \textbf{0.697} & \textbf{0.709} \\ \bottomrule
\end{tabular}
}
\caption{Retrieval performance in the M2MIR scenario. We compare the effectiveness of shifting only the documents (\textit{Doc Only}) versus shifting both queries and documents (\textit{Query and Doc}) into the unified embedding space.}
\label{tab:retrieval_mmlir}
\end{table*}

\subsection{Results}
Table~\ref{tab:retrieval_mmlir} presents the retrieval performance in the M2MIR setting.
We first observe that the \textit{Doc Only} strategy significantly improves performance over the baseline across all models. 
For instance, embeddinggemma-300m increases from 0.717 to 0.735 on average nDCG@20, and llama-nemotron-embed-1b-v2 gains from 0.515 to 0.550. 
This indicates that simply shifting the multilingual document mixture into a single English-aligned space makes the documents more accessible, even if the queries are multilingual.

However, the most robust performance is achieved with the \textit{Query and Doc} strategy.
By shifting the queries into the same English subspace as the documents, we maximize the geometric alignment between queries and relevant documents.
As shown in the results, applying \textsc{Shift} to both sides consistently yields the highest nDCG@20 and Recall@20 scores.
Notably, for embeddinggemma-300m, the average nDCG@20 further improves to 0.736, and llama-nemotron-embed-1b-v2 reaches 0.557, outperforming the \textit{Doc Only} approach.
Even for strong baselines like multilingual-e5-large and gte-Qwen2-1.5B-instruct, the \textit{Query and Doc} strategy maintains or slightly exceeds the high performance of \textit{Doc Only}.
This confirms that in M2MIR scenario, unifying both query and document representations into a common pivot space is the optimal strategy.

\section{Qualitative Analysis}
To complement our quantitative results, we present qualitative retrieval examples that illustrate how \textsc{Shift} changes the ranking behavior in practice. Each table shows English queries and the top-ranked passages retrieved by the base model and by applying \textsc{Shift}. We highlight cases where the base retriever over-prefers documents written in the query language (English), while \textsc{Shift} promotes semantically relevant passages in other languages, improving target-language exposure without introducing obvious topic drift. We report examples for four retrievers to demonstrate that these patterns are consistent across architectures and model families.

\begin{table*}[t]
\centering
\scriptsize 
\begin{tabularx}{\textwidth}{c | X c c | X c c}
\toprule
\multicolumn{7}{c}{\textbf{Q: What do some animals not have?}} \\
\midrule
\multirow{2}{*}{\textbf{Rank}} & \multicolumn{3}{c|}{\textbf{Base}} & \multicolumn{3}{c}{\textbf{SHIFT}} \\
\cmidrule(lr){2-4} \cmidrule(lr){5-7}
 & \multicolumn{1}{c}{\textbf{Content}} & \textbf{Lang} & \textbf{Relevant} & \multicolumn{1}{c}{\textbf{Content}} & \textbf{Lang} & \textbf{Relevant} \\
\midrule

1 & 
Animals are made of many cells. They eat things and digest them inside. Most animals can move. Only animals have brains (though not even all animals do; jellyfish, for example, do not have brains). Animals are found all over the earth. They dig in the ground, swim in the oceans, and fly in the sky. & 
en & 
TRUE & 
\zh{动物由许多细胞组成。动物吃了食物后会在体内将其消化。大部分动物都能移动。 只有动物有脑（但也不是所有动物都有，比如水母就没有脑）。 地球上的各个角落都有动物的身影。它们有的在地下挖洞，有的在海洋中畅游，有的在天空中翱翔。} \newline \textit{[Animals are made of many cells. They eat things and digest them inside. Most animals can move. Only animals have brains (though not even all animals do; jellyfish, for example, do not have brains). Animals are found all over the earth. They dig in the ground, swim in the oceans, and fly in the sky.]} & 
zh & 
TRUE \\
\midrule

2 & 
Out on the savanna, it is hard for a primate with a digestive system like that of humans to satisfy its amino-acid requirements from available plant resources. Moreover, failure to do so has serious consequences: growth depression, malnutrition, and ultimately death. The most readily accessible plant resources would have been the proteins accessible in leaves and legumes, but these are hard for primates like us to digest unless they are cooked. In contrast, animal foods (ants, termites, eggs) not only are easily digestible, but they provide high-quantity proteins that contain all the essential amino acids. All things considered, we should not be surprised if our own ancestors solved their ``protein problem'' in somewhat the same way that chimps on the savanna do today. & 
en & 
FALSE & 
Animals are made of many cells. They eat things and digest them inside. Most animals can move. Only animals have brains (though not even all animals do; jellyfish, for example, do not have brains). Animals are found all over the earth. They dig in the ground, swim in the oceans, and fly in the sky. & 
en & 
TRUE \\
\midrule

3 & 
Insects were the first animals to take to the air. Their ability to fly helped them evade enemies more easily and find food and mates more efficiently. Most insects have the advantage of being able to fold their wings back along the body. This gives them a wider range of small places to hide from predators. Today, the only insects that cannot fold back their wings are dragon flies and mayflies. & 
en & 
FALSE & 
\fr{Les animaux sont constitués de nombreuses cellules. Ils mangent toutes sortes de choses et les digèrent naturellement. La plupart des animaux peuvent se déplacer. Seuls les animaux ont un cerveau (même si tous les animaux n’en ont pas; les méduses, par exemple, n’ont pas de cerveau). On trouve des animaux partout sur la Terre. Ils creusent le sol, nagent dans les océans et volent dans le ciel.} \newline \textit{[Animals are made of many cells. They eat things and digest them inside. Most animals can move. Only animals have brains (though not even all animals do; jellyfish, for example, do not have brains). Animals are found all over the earth. They dig in the ground, swim in the oceans, and fly in the sky.]} & 
fr & 
TRUE \\
\midrule

4 & 
Wildlife is among the most challenging motifs for a photographer, and needs a combination of good luck, patience, experience and good equipment. Wildlife photography is often taken for granted, but like photography in general, a picture is worth a thousand words. Wildlife photography often requires a long telephoto lens, though things like a flock of birds or a tiny creature need other lenses. Many exotic animals are hard to find, and parks sometimes have rules about taking photographs for commercial purposes. Wild animals might either be shy or aggressive. The environment might be cold, hot, or otherwise hostile. & 
en & 
FALSE & 
Hewan tersusun dari banyak sel. Hewan memakan dan mencernanya. Sebagian besar hewan bisa bergerak. Hanya binatang memiliki otak (tetapi tidak semua binatang memilikinya; ubur-ubur, sebagai contoh, tidak memiliki otak). Hewan bisa ditemukan di berbagai tempat di Bumi. Hewan menggali tanah, berenang di laut, dan terbang di angkasa. \newline \textit{[Animals are made of many cells. They eat things and digest them inside. Most animals can move. Only animals have brains (though not even all animals do; jellyfish, for example, do not have brains). Animals are found all over the earth. They dig in the ground, swim in the oceans, and fly in the sky.]} & 
id & 
TRUE \\
\midrule

5 & 
Plants make oxygen which humans breathe, and they take in carbon-dioxide which humans exhale (that is, breathe out). Plants make their food from the sun by photosynthesis. They also provide shade. We make our houses from plants and make clothes from plants. Most foods that we eat are plants. Without plants, animals could not survive. & 
en & 
FALSE & 
Os animais são feitos de várias células. Eles comem coisas e as digerem em seu interior. A maior parte dos animais consegue se locomover. Apenas animais têm cérebros (embora nem todos os animais tenham; a água-viva, por exemplo, não tem cérebro). Os animais são encontrados por todo o planeta. Eles cavam o solo, nadam nos oceanos e voam pelo céu. \newline \textit{[Animals are made of many cells. They eat things and digest them inside. Most animals can move. Only animals have brains (though not even all animals do; jellyfish, for example, do not have brains). Animals are found all over the earth. They dig in the ground, swim in the oceans, and fly in the sky.]} & 
pt & 
TRUE \\
\bottomrule
\end{tabularx}
\label{tab:qualitative_analysis}
\caption{Qualitative analysis comparison on Belebele dataset. The left block shows the top retrieved documents by \textit{multilingual-e5-large}, and the right block shows the results after applying our \textsc{Shift} method.}
\end{table*}

\begin{table*}[t]
\centering
\tiny 
\begin{tabularx}{\textwidth}{c | X c c | X c c}
\toprule
\multicolumn{7}{c}{\textbf{Q: Defending democracy from foreign interference}} \\
\midrule
\multirow{2}{*}{\textbf{Rank}} & \multicolumn{3}{c|}{\textbf{Base}} & \multicolumn{3}{c}{\textbf{SHIFT}} \\
\cmidrule(lr){2-4} \cmidrule(lr){5-7}
 & \multicolumn{1}{c}{\textbf{Content}} & \textbf{Lang} & \textbf{Relevant} & \multicolumn{1}{c}{\textbf{Content}} & \textbf{Lang} & \textbf{Relevant} \\
\midrule

1 & 
Madam President, defence of democracy means enhancing transparency and it also means public participation, protecting journalists, empowering civil society, investing in education, in critical thinking and informed decision-making. It means keeping a strong framework for the rule of law and sanctioning those who ignore it. It means free voices. While we are fully aware of the importance of such a package, we need to be extremely careful on which particular instruments we choose to defend and not... & 
en & 
FALSE & 
Monsieur le Président, que dire aujourd’hui sur l’ingérence étrangère mettant en danger la démocratie? Nous nous concentrons désormais – Nous devons nous protéger et détecter les failles dans notre système qui permettent ces prises de pouvoir, que je qualifierais de souterraines. Notre ouverture sur le monde, nos libertés d’expression, de la presse et bien d’autres, ces valeurs qui nous sont si chères contiennent en elles les risques qui, parfois, nous empêchent de nous protéger. Or, nous devon... \newline \textit{[Mr. President, what can be said today about foreign interference endangering democracy? We are now focusing – We must protect ourselves and detect the flaws in our system that allow these takeovers, which I would call underground. Our openness to the world, our freedoms of expression, of the press and many others, these values which are so dear to us contain within them the risks which, sometimes, prevent us from protecting ourselves. However, we must...]} & 
fr & 
TRUE \\
\midrule

2 & 
Madam President, Vice-Presidents, colleagues, this proposal arrives at an important time as we approach the European elections in a few months and I’m very proud of our work in fighting foreign interference in the INGE and ING2 special committees in the European Parliament. Sadly, we are being proven right almost every day about how important this topic is. I believe that our special committees’ commitment has been instrumental in shaping the discourse around foreign interference, and it is enco... & 
en & 
FALSE & 
Madam President, defence of democracy means enhancing transparency and it also means public participation, protecting journalists, empowering civil society, investing in education, in critical thinking and informed decision-making. It means keeping a strong framework for the rule of law and sanctioning those who ignore it. It means free voices. While we are fully aware of the importance of such a package, we need to be extremely careful on which particular instruments we choose to defend and not... & 
en & 
FALSE \\
\midrule

3 & 
The work of the INGE Committee and this House has been a real source of inspiration for the Commission. I want to congratulate the rapporteur, Sandra Kalniete, for bringing forward this work, which seems to command broad support across this House. That is testament to the importance of the report in analysing the phenomena of foreign interference and reflecting the need for a truly whole-of-society approach. The report explores many dimensions. It looks at interference via elite capture, nationa... &
en & 
FALSE & 
Señor presidente, ¿cómo podemos defender nuestra democracia de injerencias extranjeras? Pues no permitiéndolo, no amparándolo, no legitimándolo. Porque durante demasiado tiempo hemos permitido, amparado y legitimado que Rusia e Irán, Venezuela y otros países financien e impulsen movimientos políticos extremistas en diferentes países europeos y también en esta Cámara. Porque hemos permitido, legitimado y amparado que países como Rusia apoyaran a movimientos separatistas en Europa para debilitarla... \newline \textit{[Mr. President, how can we defend our democracy from foreign interference? Well, by not allowing it, not sheltering it, not legitimizing it. Because for too long we have allowed, sheltered and legitimized Russia and Iran, Venezuela and other countries to finance and promote extremist political movements in different European countries and also in this Chamber. Because we have allowed, legitimized and sheltered countries like Russia to support separatist movements in Europe to weaken it...]} & 
es & 
TRUE \\
\midrule

4 & 
Madam President, dear Commissioners, colleagues, I have the tradition of reviewing voting results after this House votes on things the Chinese Communist Party really dislikes, and the Greek Social Democrats tend to consistently abstain or vote against China-critical texts. The underpinning problem is, of course, Greece’s critical port of Piraeus now being China-owned. This is foreign interference at work. In my country, the Netherlands, we consistently supported Nord Stream 2, the pipeline, toge... & 
en & 
FALSE & 
Comme vous le savez, la défense des principes démocratiques et la promotion d’élections libres et équitables comptent parmi les priorités de la présidence belge, et le renforcement de nos démocraties va de pair avec une autre résistance: la résistance à la désinformation, à la propagande et à l’ingérence étrangère. C’est pourquoi nous avons pris l’initiative de préparer des conclusions du Conseil sur la résilience électorale et la sauvegarde des processus démocratiques contre l’ingérence étrangè... \newline \textit{[As you know, defending democratic principles and promoting free and fair elections are among the priorities of the Belgian presidency, and strengthening our democracies goes hand in hand with another resistance: resistance to disinformation, propaganda and foreign interference. This is why we took the initiative to prepare Council conclusions on electoral resilience and safeguarding democratic processes against foreign interference...]} & 
fr & 
FALSE \\
\midrule

5 & 
Señor presidente, ¿cómo podemos defender nuestra democracia de injerencias extranjeras? Pues no permitiéndolo, no amparándolo, no legitimándolo. Porque durante demasiado tiempo hemos permitido, amparado y legitimado que Rusia e Irán, Venezuela y otros países financien e impulsen movimientos políticos extremistas en diferentes países europeos y también en esta Cámara. Porque hemos permitido, legitimado y amparado que países como Rusia apoyaran a movimientos separatistas en Europa para debilitarla... \newline \textit{[Mr. President, how can we defend our democracy from foreign interference? Well, by not allowing it, not sheltering it, not legitimizing it. Because for too long we have allowed, sheltered and legitimized Russia and Iran, Venezuela and other countries to finance and promote extremist political movements in different European countries and also in this Chamber. Because we have allowed, legitimized and sheltered countries like Russia to support separatist movements in Europe to weaken it...]} & 
es & 
TRUE & 
Madam President, Vice-Presidents, colleagues, this proposal arrives at an important time as we approach the European elections in a few months and I’m very proud of our work in fighting foreign interference in the INGE and ING2 special committees in the European Parliament. Sadly, we are being proven right almost every day about how important this topic is. I believe that our special committees’ commitment has been instrumental in shaping the discourse around foreign interference, and it is enco... & 
en & 
FALSE \\
\bottomrule
\end{tabularx}
\caption{Qualitative analysis comparison on MultiEup-v2 dataset. The left block shows the top retrieved documents by \textit{Qwen3-Embedding-0.6B}, and the right block shows the results after applying our \textsc{Shift} method.}
\label{tab:qualitative_analysis}
\end{table*}

\end{document}